\documentclass[]{aa}
\usepackage[varg]{txfonts}
\usepackage{natbib}
\usepackage{amsmath}
\usepackage{graphicx}
\usepackage{url}
\usepackage{color}
\bibpunct{(}{)}{;}{a}{}{,} 

\DeclareMathOperator{\psf}{\,psf}
\newcommand{\ew}{\mathrm{EW_{CaT}}}

\title{Resolving stellar populations with crowded field 3D spectroscopy\thanks{Based on observations collected at the Centro Astron\'{o}mico Hispano Alem\'{a}n (CAHA) at Calar Alto, operated jointly by the Max-Planck Institut f\"ur Astronomie and the Instituto de Astrof\'{\i}sica de Andaluc\'{\i}a (CSIC)}\fnmsep\thanks{Based on observations made with the NASA/ESA Hubble Space Telescope, and obtained from the Hubble Legacy Archive, which is a collaboration between the Space Telescope Science Institute (STScI/NASA), the Space Telescope European Coordinating Facility (ST-ECF/ESA) and the Canadian Astronomy Data Centre (CADC/NRC/CSA).}}
\author{S.~Kamann
  \and L.~Wisotzki
  \and M.~M.~Roth}

\institute{Leibniz-Institut f\"ur Astrophysik Potsdam (AIP), An der Sternwarte 16, 14482 Potsdam, Germany\\ \email{skamann@aip.de}}

\date{Received / Accepted}

\abstract{
We describe a new method to extract spectra of stars from observations of crowded stellar fields with integral field spectroscopy (IFS). Our approach extends the well-established concept of \emph{crowded field photometry} in images into the domain of 3-dimensional spectroscopic datacubes. The main features of our algorithm are: (1) We assume that a high-fidelity input source catalogue already exists, e.g.\ from HST data, and that it is not needed to perform sophisticated source detection in the IFS data. (2) Source positions and properties of the point spread function (PSF) vary smoothly between spectral layers of the datacube, and these variations can be described by simple fitting functions. (3) The shape of the PSF can be adequately described by an analytical function. Even without isolated PSF calibrator stars we can therefore estimate the PSF by a model fit to the full ensemble of stars visible within the field of view. (4) By using sparse matrices to describe the sources, the problem of extracting the spectra of many stars simultaneously becomes computationally tractable. We present extensive performance and validation tests of our algorithm using realistic simulated datacubes that closely reproduce actual IFS observations of the central regions of Galactic globular clusters. We investigate the quality of the extracted spectra under the effects of crowding with respect to the resulting signal-to-noise ratios (S/N) and any possible changes in the continuum level, as well as with respect to absorption line spectral parameters, radial velocities and equivalent widths. The main effect of blending between two nearby stars is a decrease in the S/N in their spectra. The effect increases with the crowding in the field in a way that the maximum number of stars with useful spectra is always $\sim$0.2 per spatial resolution element. This balance breaks down when exceeding a total source density of $\sim$1 significantly detected star per resolution element. We also explore the effects of PSF mismatch and other systematics. We close with an outlook by applying our method to a simulated globular cluster observation with the upcoming MUSE instrument at the ESO-VLT.
}


\begin{document}

\maketitle

\section{Introduction}
\label{sec:intro}

The observation of spatially resolved, but densely populated stellar fields such as star clusters or the inner regions of very nearby galaxies poses a severe challenge because of crowding. At finite angular resolution each star is represented by the point spread function (PSF) characteristic of the system. Observable is only the superposition of many scaled and shifted PSFs, according to brightness and position of the stars in the field of view. It is then a major task to disentangle the various overlapping PSFs in order to measure the corresponding stellar fluxes as accurately as possible.

In imaging photometry, this challenge was addressed already many years ago, and dedicated analysis tools have been developed to perform what is commonly termed `crowded field photometry'. \texttt{daophot} \citep{1987PASP...99..191S} is the most widely used of such tools and has been applied extensively and with great success. At the core of \texttt{daophot} (and similar software) is a code that fits a PSF model simultaneously or in succession to all stars within a given field. The resulting photometry is unbiased with respect to overlapping stellar images (within the accuracy of the PSF model and as long as the list of fitted stars is complete), so that even heavily blended stars can be recovered.

However, the astrophysically desirable move from photometry to spectroscopy is still fraught with difficulties when it comes to crowded stellar fields. Traditional long-slit (or multi-slit) spectroscopy is clearly limited by the amount of blending within the slit aperture, which under conditions of heavy crowding may be almost impossible to control. The same is true for fibre-fed multiplexing spectroscopy using single apertures. Consequently, investigations using this type of equipment have typically restricted themselves to regions with modest crowding, such as the outer parts of star clusters or galaxies, and/or focused on the brightest sources in the fields of interest. 
A notable exception is the study by \citet{2002AJ....124.3255V} and \citet{2003AJ....125..376G} who performed HST long-slit spectroscopy of the center of the globular cluster M15.
Here the improved spatial resolution of HST helped in dealing with the limitations of traditional spectroscopy.

Potentially much more powerful in this domain is the direct combination of imaging and spectroscopy. `Integral Field Spectroscopy' (IFS) -- often also called 3D or IFU spectroscopy (IFU = integral field unit) --  has matured into a widely used observing technique over the last decade. It is particularly powerful for extended objects such as galaxies, providing access to spectroscopic information across the full spatial extent of the target. Integral field spectroscopy has strongly advanced our knowledge of galaxies, both in the local universe \citep[e.g.,][]{2002MNRAS.329..513D,2011MNRAS.413..813C} and at higher redshifts \citep[e.g.,][]{2009ApJ...706.1364F}.

Beyond the ability to trace the variation of observable properties across the field of view (FoV), IFS data also provide the user with an unprecedented access to single out individual objects surrounded by (and possibly blended with) other sources. This can be done in increasing levels of sophistication. A straightforward approach in stellar fields is to identify spatial pixels that are dominated by the contribution of a single star and obtain a spectrum of that star by summing the spectra in only those pixels. An example is the recent work by \citet{2011A&A...530A.108E}, who used IFS to obtain spectra of massive stars in the Tarantula nebula (30 Doradus). A similar treatment was applied to near-infrared IFS observations of stars in the Galactic centre by \citet{2005ApJ...628..246E} and \citet{2011ApJ...741..108P}. While scientifically very successful, this approach is obviously still limited to moderate crowding conditions.

In order to facilitate the extraction of the spectra, knowledge of the spatial shape of the sources must be incorporated. Especially when dealing with barely resolved or yet unresolved sources, a proper knowledge of the PSF becomes extremely important. Unfortunately, the fields of view of existing IFS instruments are so small that they often do not contain any undisturbed PSF calibrator. In order to proceed it thus becomes necessary to estimate the PSF directly from the data. Several authors have presented solutions of this task, although usually restricted to specific science applications. For the severely background-limited spectrophotometry of planetary nebulae (PN) in the bulge of M31, \citet{2004ApJ...603..531R} estimated the PSF from a bright PN image in [\ion{O}{iii}] and fitted this to each wavelength bin of the datacube. \citet{2005A&A...437..217F} applied the \texttt{cplucy} algorithm \citep{1994rhis.conf...79L} with input from high spatial resolution HST images to extract spectra of LBV candidates in M33, while fully accounting for a highly variable unresolved background. Technical details of these two applications are discussed in \citet{2004AN....325..155B}. In a similar spirit, the bright nuclear emission lines of quasars can be used to `self-calibrate' the PSF for the analysis of quasar host galaxies \citep{2004AN....325..128J,2006A&A...459..717C,2011A&A...535A..72H}. However, due to the lack of emission line point sources, these procedures cannot be applied to arbitrary stellar fields. Another limitation lies in the fact that emission lines provide the PSF only at specific wavelengths, and an extrapolation to the full spectral range may be doubtful.

A more general approach was explored by \citet{2003A&A...408..455W} who deblended the four overlapping components of the gravitationally lensed QSO HE~0435-1223 from IFS data by assuming a purely analytic (but wavelength-dependent) PSF shape that was iteratively optimised. The main limitation of this particular solution was its computational inefficiency, which would make an application to fields with many more sources prohibitive. \citet{2007MNRAS.376..125S} extended this approach to non-point sources in an IFS study of the galaxy cluster Abell 2218, where they used the morphological information on individual galaxies obtained from high-resolution imaging to deblend the overlapping data into individual galaxy spectra. 

In this paper we present a new algorithm for the spectrophotometric analysis of generic crowded stellar fields observed by integral field spectroscopy. The algorithm determines a fully self-calibrated wavelength-dependent PSF model which is subsequently fitted to the entire datacube. In many aspects the algorithm is an extension of the well-established `crowded field photometry' approach, but the spectroscopic dimension requires the addition of some genuinely new features. Note that while we were completing this paper, \citet{2012A&A...540A..48S} published an analysis of IFS observations in the Galactic bulge which includes PSF estimation and bears some generic resemblance to our approach. Unfortunately, not many details are provided in that paper on the algorithm itself and its performance.

The purposes of the present paper are twofold. Firstly, the paper provides the methodical foundations for a number of subsequent articles focusing on the central regions of globular clusters (Kamann et al., in prep.). Here we use some of those data for illustration purposes only. Secondly, we also want to highlight the huge potential of IFS for crowded field spectroscopy in general, especially in view of the upcoming wide-field panoramic integral-field spectrograph MUSE at the ESO-VLT.

The paper is organized as follows. In Sect.~\ref{sec:idea} we start with an exposition of the basic considerations that went into the development of our method. An overview of the observational data that motivated this study, and of the simulated data that we created to test our algorithm, is given in Sect.~\ref{sec:data}. The analysis algorithm itself is presented in detail in Sect.~\ref{sec:algorithm}. Section~\ref{sec:performance} describes the extensive tests that we have carried out to validate its performance. A discussion of potential sources that could systematically influence the performance of the algorithm is presented in Sect.~\ref{sec:systematics}. In Sect.~\ref{sec:muse} we demonstrate briefly the application of our method to data that will soon be obtained with the MUSE instrument. We wrap up our conclusions in Sect.~\ref{sec:conclusions}.

\section{Basic concepts}
\label{sec:idea}

In a simplified picture, an integral field datacube can be considered a sequence of monochromatic images, hereafter called layers. A straightforward approach would then be to use existing methods to analyze the data cube layer by layer, i.e. to perform crowded field \emph{photometry} individually on each layer, and combine the results afterwards. Yet, as we shall demonstrate in the following, such an approach does not use the full potential of IFS data. 

The development of our new algorithm was guided by the following thoughts:

(1) For many objects, like most globular clusters or several nearby galaxies, high resolution images are already available \citep[e.g., the HST/ACS survey of galactic Globular Clusters by][]{2007AJ....133.1658S}. The depth reached by these images is usually sufficient to assume that all stars that can be resolved with a (seeing-limited) IFU have been detected, and that their relative positions are known. Thus, an inventory of stars in the observed field already exists, and there is no need to perform sophisticated source detection on the integral field data. But of course the question remains which sources from an available catalogue can be recovered in the datacube. Given the typically lower spatial resolution of IFS data, only a subset of the catalogued sources will be accessible to the analysis. We describe in Sect.~\ref{sec:selection} how we derive an optimal subset of sources from a catalogue that can be extracted reliably.

(2) The individual monochromatic layers of a data cube are not independent from one another. Because all of them were observed simultaneously, temporal effects like seeing variability, atmospheric dispersion and instrument flexure affect all images in a way that properties such as the PSF or source positions are interconnected. While they may vary from layer to layer, these variations will be smooth with wavelength. We thus should use the whole datacube to determine one (wavelength dependent) PSF model and one set of (wavelength dependent) coordinates.
The potential of IFS to obtain coordinates with a precision beyond that achievable with a single image at the same spatial resolution has been demonstrated previously \citep[e.g.,][]{1998ApJ...503L..27M,2003A&A...408..455W}.

(3) For the current generation of integral field spectrographs, the relatively small field of view often implies that sufficiently bright and isolated PSF calibrator stars are not available within a science exposure. Dedicated (i.e.\ non-simultaneous) observations of isolated stars are of very limited use for accurate PSF modelling  because of the temporal variations in the PSF. In the future it may become possible to predict the PSF from adaptive optics wavefront sensor data (e.g., \citealt{2010SPIE.7736E..48J}). For the present, however, we are bound to estimate the PSF directly from the crowded field science data. Nevertheless, this information is actually there -- not in individual stars, but encased in the ensemble of all stars in the field. In Sect.~\ref{sec:psf} we present an iterative procedure to construct a global PSF model by simultaneous fitting of the full ensemble.

(4) Besides the spectra of individually resolved stars, an IFS data cube may also contain a background of fainter and possibly unresolved stars, and/or nebular emission. This quasi-diffuse component can be a major nuisance and source of systematic errors during the extraction of stellar spectra, but it is also possible that the scientific focus is mainly directed towards this component. For these reasons we made a special effort to find an appropriate solution to account for such a non-trivial background component in the analysis. This is explained in detail in Sect.\ref{sec:algorithm}.

\section{Data}
\label{sec:data}

\subsection{Observations of globular clusters}
\label{sec:obs}

\begin{figure*}[tb]
 \centering
 \includegraphics[width=17cm]{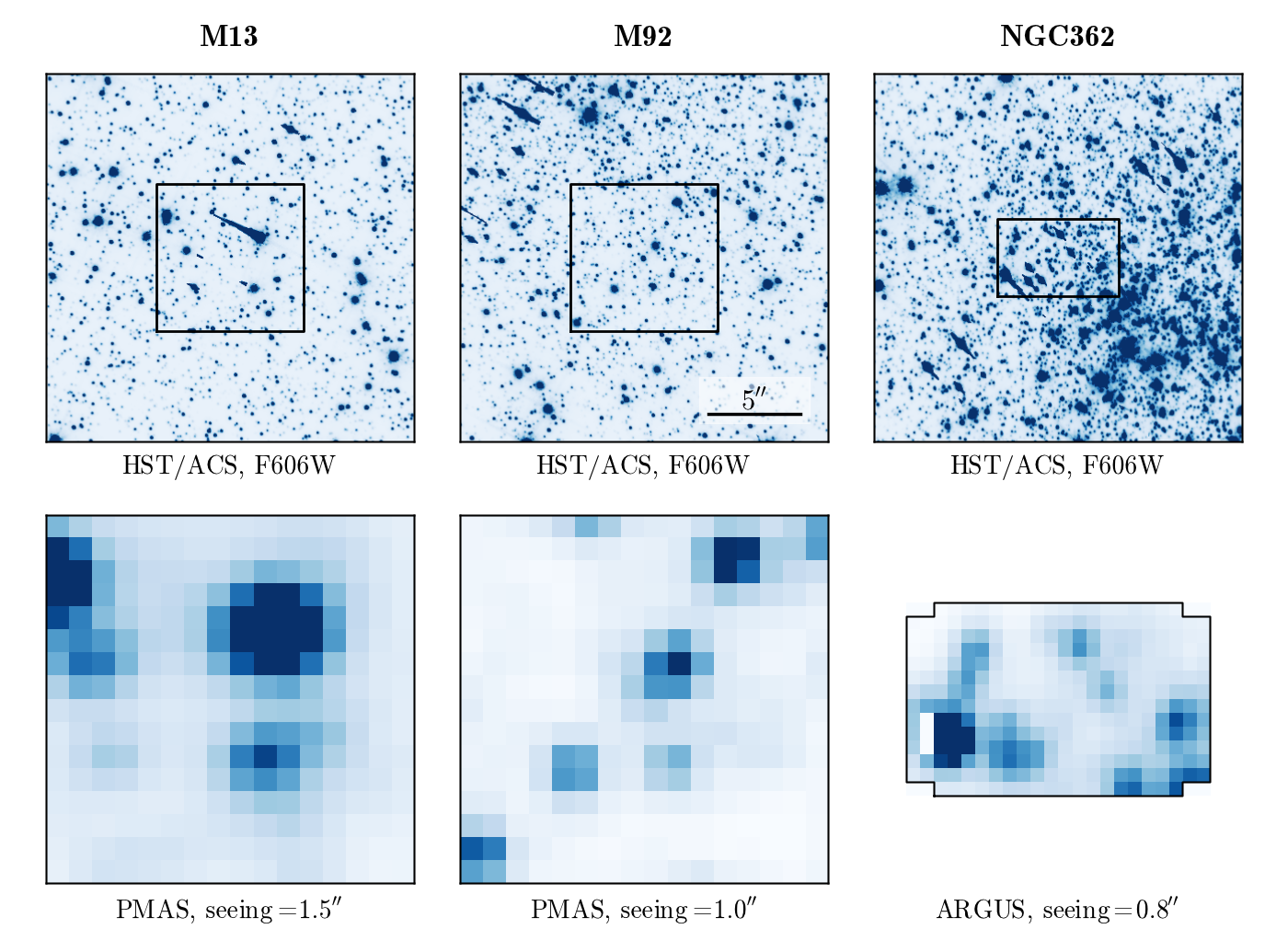}
 \caption{Impressions from the observations of several globular clusters that we obtained using the integral field spectrographs PMAS and ARGUS. The upper row shows a $20\arcsec\times20\arcsec$ cut-out from an HST/ACS image with the footprint of a single integral field observation. A whitelight image of each integral field data cube is shown below. In each image, north is up  and east is left.}
 \label{fig:data_overview}
\end{figure*}

We recently collected 3D spectroscopy data of a sample of Galactic Globular Clusters with the instruments PMAS \citep{2005PASP..117..620R} and ARGUS \citep{2002Msngr.110....1P}. In each cluster we mapped the central region out to a distance comparable to the core radius of the cluster. The main scientific aims are (1) a search for binary stars, and (2) to place constraints on the possible presence of intermediate-mass black holes in the clusters. The results from these observations will be presented in forthcoming papers (Kamann et al, in prep.). 

The instrument setups in these observations varied a little from campaign to campaign, but were generally chosen to facilitate a kinematic analysis given the expected velocity dispersions of the clusters. The typical spectral resolution $R = \lambda/\Delta\lambda$ was $\sim$8000, and the spectral range was targeted at the Calcium triplet ($\lambda\lambda$8498~\AA, 8542~\AA, 8662~\AA). In order to sample the PSF in the best possible way we always selected the smallest available spaxel scale (spaxel = spatial pixel), 0\farcs3 for ARGUS and 0\farcs5 for PMAS. With $16\times16$ spaxels for PMAS and $22\times14$ for ARGUS we thus cover an area per pointing of $8\arcsec\times8\arcsec$ and $6\farcs6\times4\farcs2$, respectively. The seeing conditions were variable, with $\sim$0\farcs7 in the best and 1\farcs4 in the worst case. Some whitelight images are presented Fig.~\ref{fig:data_overview} to give an impression of the data. The crucial dependence of the data quality on the seeing during the observations is clearly visible.

\subsection{Simulated data}
\label{sec:sim}

\begin{figure*}[tb]
  \centering
 \includegraphics[width=17cm]{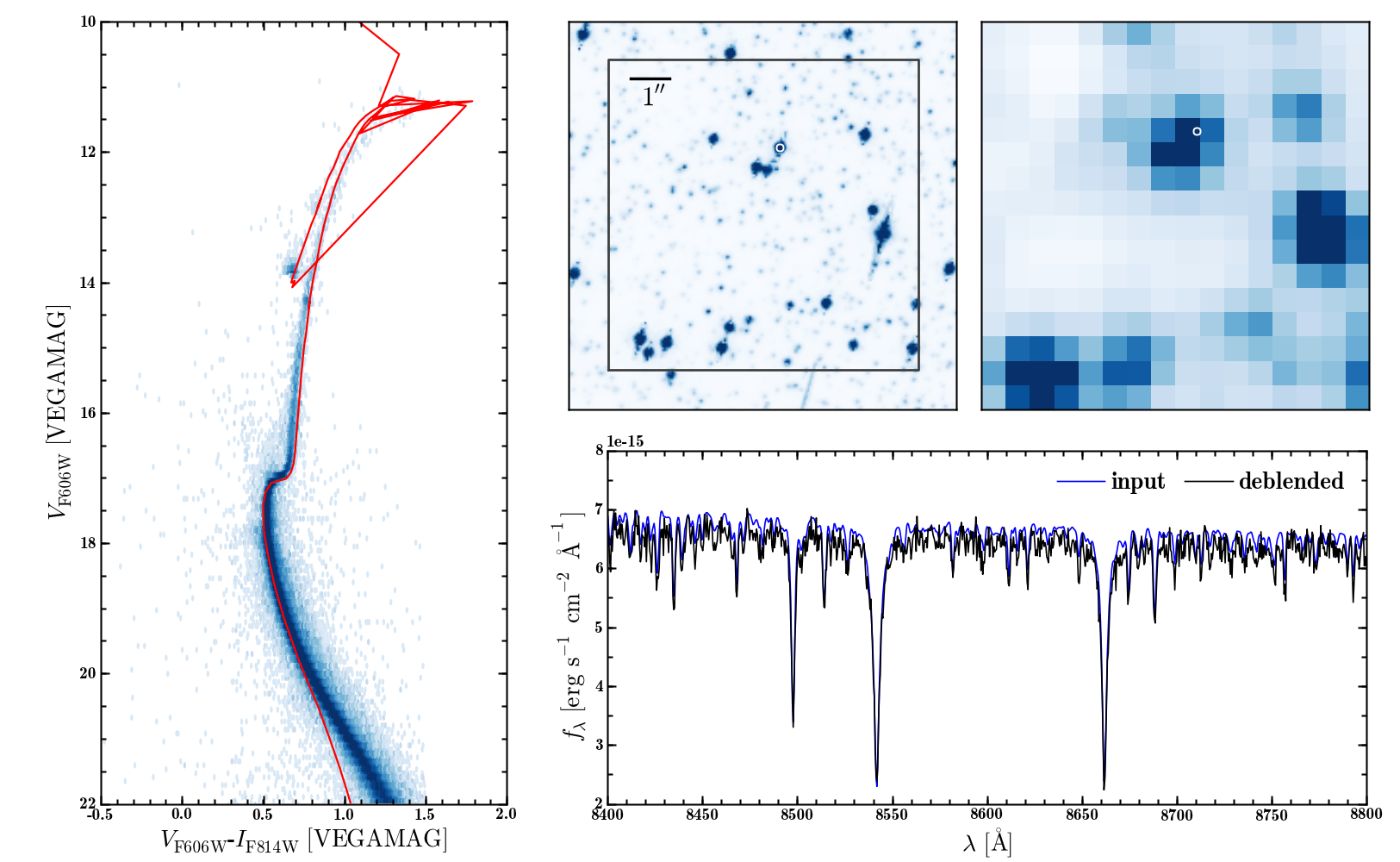}
 \caption{Illustration of how the simulated data was produced and processed. Left panel: Input colour-magnitude diagram of 47Tuc. Overplotted is the isochrone that was used to assign spectra to the individual stars. Upper panels: Cut-out from an HST image of 47Tuc (upper middle); whitelight image of a simulated crowded field datacube centred on the HST footprint (upper right). The lower right panel shows the deblended spectrum of the star marked with a white circle. The input spectrum of this star is overplotted.}
 \label{fig:sim_catrip_cube_prep}
\end{figure*}

To assess the the performance of our algorithm and to validate the quality of the deblended spectra we carried out extensive simulations. In line with our main scientific interest we focus on Globular Clusters, but the results are of course applicable to any sort of crowded stellar fields. At any rate the central regions of Globular Clusters display some of the most challenging cases of crowding that current instruments are capable of dealing with.

The preparation of the simulated data can be summarized as follows: We used the $V$- and $I$ band photometry of the globular cluster 47Tuc from \citet{2007AJ....133.1658S} and \citet{2008AJ....135.2055A}. To assign a realistic spectrum to each star, we constructed a single isochrone ($t=13\:\mathrm{Gyr}$, $Z=0.0045$), using the tool of \citet{2008A&A...482..883M}%
\footnote{available at \url{http://stev.oapd.inaf.it/cgi-bin/cmd_2.3}}. For each star we thus obtained estimates of $T_\mathrm{eff}$ and $\log g$ that were in turn used to select an appropriate spectrum from the stellar library of \citet{2005A&A...442.1127M}. The library spectra have an intrinsic spectral resolution of $R=20000$ and cover the entire wavelength range from 2500~\AA$\,$ to 10000~\AA. In order to resemble our observational data, we extracted only the region around the Calcium triplet and convolved the spectra with a Gaussian to obtain a final resolution of $R\sim7000$. 

To simulate realistic datacubes, random fields of $8\arcsec\times8\arcsec$ (the FoV of the PMAS instrument) were selected from the central region of 47Tuc. Stars in those fields were represented by their respective spectra, and a velocity drawn randomly from a normal distribution with $\sigma=10\mathrm{km/s}$, similar to the velocity dispersion in the centre of this cluster \citep{2006ApJS..166..249M}, was assigned to each spectrum. Using an analytical PSF profile with a full width at half maximum (FWHM) corresponding to a seeing around $1.0\arcsec$, cubes with $16\times16$ spatial pixels were prepared. Finally, appropriate noise was added. The resulting spatial sampling was $0.5\arcsec/{\rm pixel}$, again to resemble observational data obtained with PMAS. An example of the simulated data is given in Fig.~\ref{fig:sim_catrip_cube_prep}. In the following, we refer to those datacubes as the \emph{`crowded field datacubes'}.

Besides these rather realistic datacubes, we also prepared idealized cubes containing only two stars. We used these to investigate different aspects of the deblending procedure in isolation. The spectra of the stars were assigned using the same procedure as for the crowded field datacubes. We refer to these simulated data as the \textit{`two star datacubes'}.

\section{Deblending and extraction of spectra: Algorithm}
\label{sec:algorithm}

\subsection{Global model}
\label{sec:fitting}

We first clarify our adopted symbol and naming scheme. The datacube at hand is supposed to have pixel values $\mathfrak{d}_{i,j,k}$, where the first two indices relate to spatial coordinates and the index $k$ to the spectral axis. To denote different stellar sources in the datacube we use the superscript $n$ while background components have a superscript $m$. A model datacube is then described as the sum of all sources that contribute flux,

\begin{equation} \label{eq:model}
 \mathfrak{m}_{i,j,k} = \sum_n f^n_k \psf^n_{i,j,k} + \sum_m b^m_{i,j,k}\:.
\end{equation}

In Eq.~\ref{eq:model}, $f^n_k$ is the total monochromatic flux of star $n$ in layer $k$, and $\psf^n_{i,j,k}$ is the fraction of $f^n_k$ in the considered pixel. This fraction is equal to the value of the normalized PSF of star $n$ at the spatial position $(i,j)$ and at spectral pixel $k$.
The flux in each pixel of background component $m$ is given by $b^m_{i,j,k}$. For a spatially constant background (such as the night sky), $b^m_{i,j,k}$ will only depend on $k$. An example for a background whose intensity varies across the field of view is the contribution of  unresolved stars to the observed flux; this will be discussed in Sect.\ref{sec:background}. The second sum in Eq.~\ref{eq:model} can be further expanded to also include spatially resolved sources that might be present in the datacube. An example for such a component might be gaseous line emission in the observation of a young star cluster. Here we restrict the discussion to fields in where the flux is dominated by stellar continuum sources.

Starting from Eq.~\ref{eq:model}, we then search for the best (in a least-squares sense) model for a given dataset. We therefore have to minimize

\begin{equation} \label{eq:chi2}
 \chi^2 = \sum_{i,j,k} \frac{\left( \mathfrak{d}_{i,j,k} - \sum_n f^n_k \psf^n_{i,j,k} - \sum_m b^m_{i,j,k} \right)^2}{\sigma_{i,j,k}^2}  \:.
\end{equation}

Here, $\sigma_{i,j,k}^2$ is the variance tailored to each pixel value. Unfortunately, straightforward minimization of Eq.~\ref{eq:chi2} is computationally very demanding for the large parameter space that needs to be covered: in each layer $k$, a stellar source will contribute $3$ free parameters ($f^n_k$, $x^n_k$ and $y^n_k$) and a background component will contribute at least one additional free parameter. In addition, several free parameters might be needed to find a suitable model for the PSF of the observation. A further complication is that Eq.~\ref{eq:chi2} represents a non-linear minimization problem.

In order to make the search for a solution feasible, we split the optimization into three tasks: (i) an optimization for the PSF, (ii) an optimization for the source coordinates, and (iii) an optimization for the fluxes. In each step, the model is optimized only with respect to one of these three properties, while the other two remain fixed to their current value. After one step has converged, the model parameters currently in focus are updated to their best-fit values and the model is optimized for the next set of parameters. On each layer $k$, the steps (i) to (iii) are then iteratively repeated until convergence is found for the fluxes of the sources.

The practical implementation of our approach can be summarized as follows: we start with an initial guess for the PSF and the source coordinates and use them to fit the fluxes in the central layer of the datacube. The reason for starting at the central layer is that the efficiency of the spectrograph should be highest at the center of the covered wavelength range and therefore the data should give the tightest constraints on the model. After the iteration has converged to a solution for the central layer, we continue with the two adjacent layers. In this way, the analysis proceeds simultaneously to the red and blue end of the covered wavelength range. An integral field datacube provides a convenient structure for such an iterative approach, as the changes in the PSF or source coordinates between two adjacent layers are always small. So the best-fit model of the previous layer is already a very good starting point for the analysis of the next one. A potential drawback of this approach is that if a single layer returns a strongly deviating model, such caused by an undetected cosmic-ray hit or strong telluric absorption, it will affect the analyses of following layers. To avoid this, we generate initial guesses by averaging over the best-fit models of the last $N$ layers, with $N$ typically being of the order of $10$.

In general, we do not consider cosmic ray hits as a major problem. The structure of raw IFS data actually allows for a convenient method to remove cosmic rays before performing the data reduction \citep{2012A&A...545A.137H}. Additionally, it is possible to significantly reduce the influence of undetected cosmic rays in the analysis. \citet{1987PASP...99..191S} presented a robust scheme that dynamically reduces the weights of pixels with strong residuals that is also applicable to IFS data.

After all layers of the cube have been processed, their respective best-fit models are combined to produce a coherent wavelength-dependent PSF model and to determine source coordinates as a smooth function of wavelength. Using this information, the datacube is then processed for a second time, yet this time only the fluxes are fitted and we obtain the final spectra of the sources.

Note that after a few layers have been analysed, the analyses of the red and blue half of the datacube proceed independently from each other. Thus, comparing the results obtained at the red and blue end can be used to check the reliability of the obtained model.

\subsection{Modelling the PSF}
\label{sec:psf}

The usual approach to determine the PSF in crowded field photometry is to select a number of relatively isolated stars and fit them with an analytical function. To account for possible mismatches between the analytical profile and the shape of the true PSF, an empirical look-up table correction is frequently applied afterwards.

As discussed above, this approach cannot be applied to IFS datacubes without modification, mainly because of the very small FoV. Our adopted approach is to instead use the full ensemble of resolved stars within a field to reconstruct a global PSF model, obtained by means of a least squares fit of Eq.~\ref{eq:chi2} to the data. The approach to recover the PSF using all stars in the field has previously been used by \citet{1993PASP..105.1342S}. A notable difference of our implementation is that the PSF is fitted to all stars simultaneously instead of in a sequential manner. Similar to \citet{1993PASP..105.1342S}, we restrict the PSF description to a purely analytical model, since the construction of a reliable look-up table not only requires sufficiently isolated stars, but also a very well-sampled PSF. With the somewhat coarse spatial sampling of many IFUs this cannot always be taken for granted. But as the typical signal-to-noise ratio (S/N) per spectral pixel is much lower than in broad-band images, an analytical model should be adequate in most cases.

To define an analytical PSF model we follow the approach adopted in \texttt{GALFIT} \citep{2002AJ....124..266P}. In general, the PSF will not be round but rather of elliptical shape, with ellipticity $e$ and position angle $\theta$. To account for the ellipticity, the pixel coordinates $x$ and $y$ are first transformed into a coordinate system $({\hat x},\, {\hat y})$ centered on the origin of the star and whose ${\hat x}$-axis is aligned with the semi-major axis of the PSF:
\begin{equation} \label{eq:psf_x}
 {\hat x} = (x-x^n)\cos\theta - (y-y^n)\sin\theta ,
\end{equation}
\begin{equation} \label{eq:psf_y}
 {\hat y} = (x-x^n)\sin\theta + (y-y^n)\cos\theta ,
\end{equation}
with $(x^n, y^n)$ being the pixel coordinates of source $n$.
The distance to the centre of the PSF can now be written as as an angle-dependent quantity,
\begin{equation} \label{eq:psf_radius}
 r(x, y) = \sqrt{{\hat x}^2 + \left( \frac{\hat y}{1-e} \right)^2} \, .
\end{equation}
In order to describe the radial shape of the PSF we adopt the Moffat profile, with a functional form given as
\begin{equation} \label{eq:psf_moffat}
 M(x, y) = \Sigma_\text{0} \left(1 + \left(\frac{r(x,y)}{r_\text{d}}\right)^2\right)^{-\beta} \, .
\end{equation}
The width of the Moffat profile is mainly determined by the dispersion radius $r_\text{d}$, while the $\beta$-parameter defines the kurtosis of the profile, i.e the broadness of the wings of the PSF. The FWHM of the Moffat profile can be expressed in terms of $r_\text{d}$ and $\beta$ as
\begin{equation} \label{eq:psf_moffat_fwhm}
 \mathrm{FWHM} = 2 \sqrt{2^{1/\beta}-1}\,r_\mathrm{d} \, .
\end{equation}
This leaves us with 4 free PSF shape parameters per layer: $\beta$, FWHM, $e$, and $\theta$. Depending on the quality of the data to be analyzed, the number of free parameters can be reduced, for example by assuming a Gaussian instead of a Moffat profile, or by enforcing a circular PSF.
The central intensity $\Sigma_\text{0}$ of the PSF is directly tied to the monochromatic flux of a source.

\begin{figure}[tb]
  \resizebox{\hsize}{!}{\includegraphics{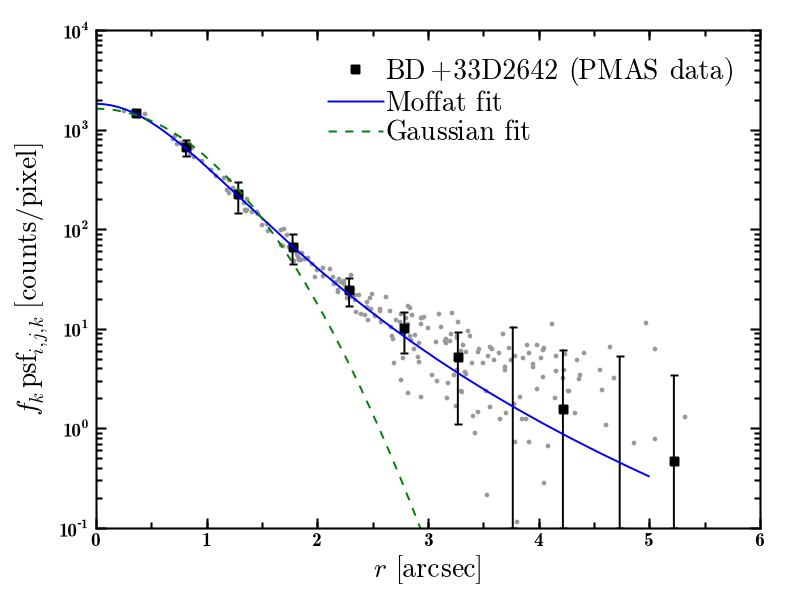}}
  \caption{Radial surface brightness profile of the star BD+33D2642 as measured in a PMAS datacube.
Small grey dots show the flux in individual pixels, the black squares were obtained by radially binning the data.
We used a Moffat profile and a simple Gaussian profile as the analytical PSF description and fitted the star with both PSF models.
The blue solid line shows the best-fit PSF when using the Moffat, the green dashed line shows the best-fit PSF for the Gaussian profile.}
 \label{fig:psf_fit_stdprof}
\end{figure}

For each star $n$ in the analysis, an empty array of the same size as a layer of the datacube is created. Its pixel values are set to the intensity of a normalized PSF at a radius $r(x,y)$, using the current best-fit values for the source coordinates ($x^n_k$, $y^n_k$). Yet directly using Eq.~\ref{eq:psf_moffat} to determine the pixel values $\psf^n_{i,j,k}$ can lead to systematic errors close to the origin of the profile if the sampling of the PSF approaches critical values. In such cases, the change of $M(x,y)$ within one pixel is so strong that its value at the centre of a pixel is not a good approximation for the integrated PSF intensity in that pixel. Currently, we solve this problem by supersampling each pixel within a certain distance to the centre of the PSF by a factor of typically 25--100 and calculate the intensity for each subpixel. The final value of each pixel then is obtained by summing over its subpixels. Numerical integration schemes such as the one presented by \citet{1989PASP..101..294B} can significantly reduce the number of required subpixels and will be considered in our ongoing development of the algorithm. Once a PSF has been prepared for each source, we can use Eq.\ref{eq:chi2} to optimize the PSF model for each layer of the datacube. After all layers have been processed, the results of the individual layers are modelled as a smooth function of wavelength for each free parameter of the PSF model. Following \citet{2003A&A...408..455W}, we use low order polynomials for this task. This way we finally obtain the wavelength dependent PSF.

To illustrate that our approach results in a valid description of the PSF in integral field data we analyzed one of our PMAS datacubes of the standard star BD+33D2642. The datacube contains a single star and thus allows for a precise measurement of the PSF even in the faint wings. We analysed the data in the way just described, using a single point source and a background component. In Fig.~\ref{fig:psf_fit_stdprof} a comparison between a PSF as it is measured from our PMAS data and our analytical profiles is shown. It is obvious that a Moffat profile provides a good overall representation of the PSF. On the other hand, using a Gaussian severely underestimates the wings of the PSF.

\begin{figure}[tb]
  \resizebox{\hsize}{!}{\includegraphics{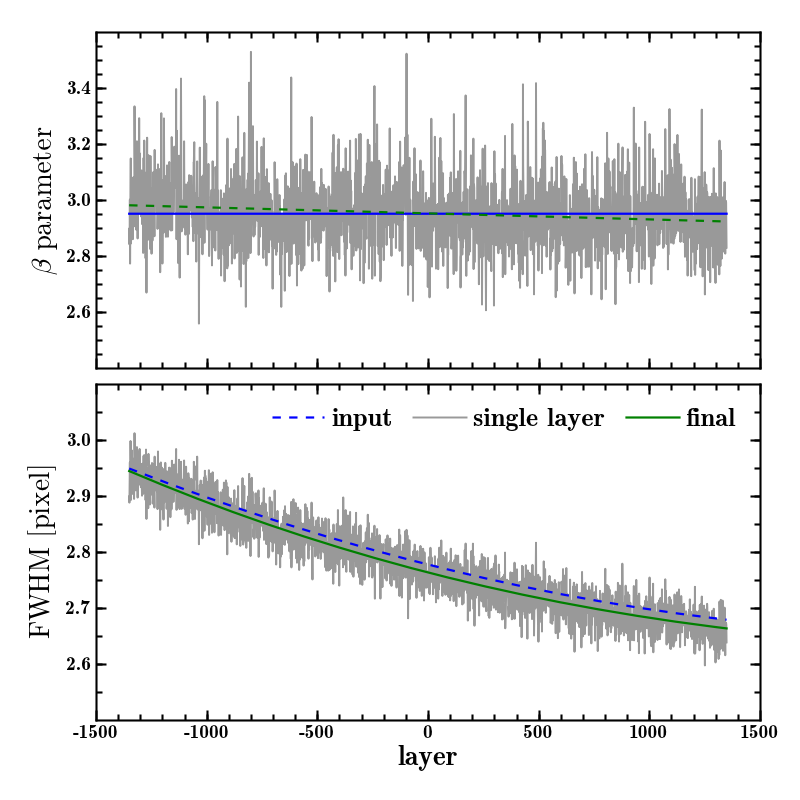}}
  \caption{Example for the recovery of the PSF in a datacube.
Shown is the accuracy of the recovered PSF parameters $\beta$ (\textit{top}) and FWHM (\textit{bottom}) in comparison to their true values (blue dashed line).
The grey solid line gives the best-fit values of the PSF parameters obtained in each layer and the final parameters of the PSF model obtained from these values are shown as a solid green line.
Note that the analysis was started at the central image of the datacube and proceeded simultaneously to the red and blue end of the cube.}
\label{fig:psf_accuracy_single}
\end{figure}

\begin{figure*}[tb]
 \centering
  \includegraphics[width=17cm]{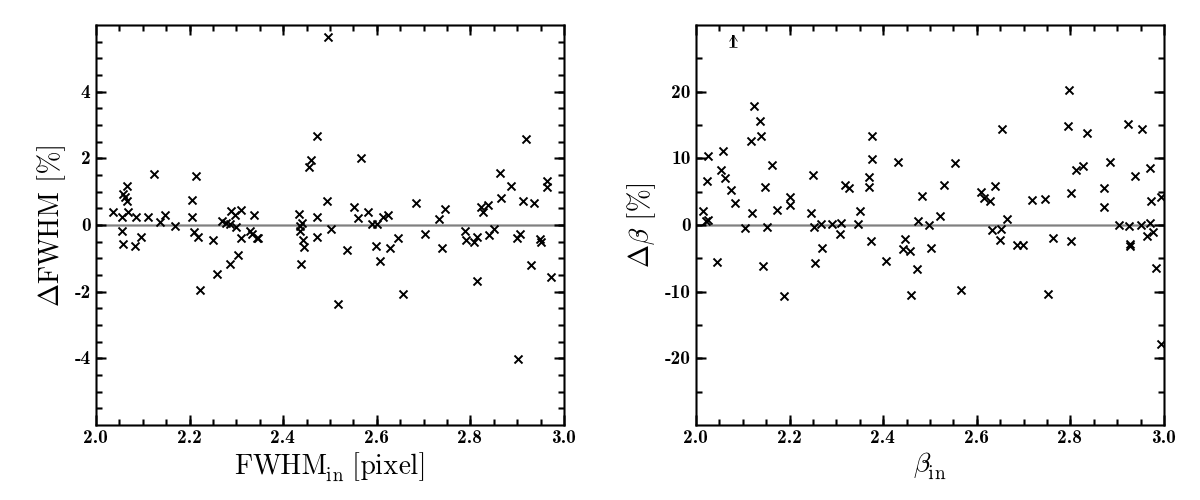}
  \caption{Comparison between recovered and input PSF parameters for 100 simulated crowded field datacubes. The PSF was modelled using a Moffat profile with wavelength-dependent FWHM and constant $\beta$ parameter. Shown is the mean fractional difference in percent, between recovered and true FWHM (\textit{left}) and $\beta$ parameter (\textit{right}), respectively, as a function of the input values. A single outlier of $\Delta\beta\sim60\%$ falls out of the plotted range and is shown as a lower limit.}
  \label{fig:psf_accuracy_statistics}
\end{figure*}

We used the simulated `crowded field datacubes' to investigate how well our approach can recover the PSF in a dense stellar field. We produced 100 cubes based on a wavelength-dependent PSF represented by a Moffat profile with constant $\beta$ and varying FWHM. An example for the PSF recovery is shown in Fig.~\ref{fig:psf_accuracy_single}. It demonstrates that the constraints on the PSF in an individual layer are not very stringent, as can be seen by the large scatter especially for the $\beta$ fits. Yet the final combination of all individual fits into a wavelength-dependent model yields a very close representation of the true PSF.

To obtain a statistical impression on the accuracy of the PSF recovery we analysed the full sample of 100 simulated `crowded field datacubes' and compared the recovered PSF parameters to the input. The results of this comparison are shown in Fig.~\ref{fig:psf_accuracy_statistics}. 

These plots show that the FWHM of the PSF can usually be recovered to an accuracy of $1$--$2\%$, while the recovered value of $\beta$ usually agrees with the input value to within $10\%$. Thus, the PSF width is better constrained than the $\beta$ parameter, similar to what we observed in Fig.~\ref{fig:psf_accuracy_single}. This is well known and due to the fact that $\beta$ governs the outer wings of the PSF where the signal is generally low. But this also implies that errors in $\beta$ do not influence the overall optimisation (Eq.~\ref{eq:chi2}) as much as mismatches in FWHM. 

Furthermore, it is reassuring that the recovered values of the FWHM are unbiased in the sense that they scatter around the `true' value. This is not strictly the case for $\beta$, which tends to come out slightly too high (by a few percent), thus corresponding to a PSF with slightly less pronounced wings than was put into the simulations. This is probably due to the fact that for high source densities, the ensemble of faint wings of each PSF can be misidentified as being part of the background; so apparently a very small fraction is, on average, transferred out of the wings of the stars used for the PSF estimation and into the background component. We discuss the consequences of the achievable PSF accuracy on the quality of the extracted spectra in detail in Sect.~\ref{sec:performance}.

\subsection{Source positions}
\label{sec:positions}

\begin{figure*}[tb]
 \centering
  \includegraphics[width=17cm]{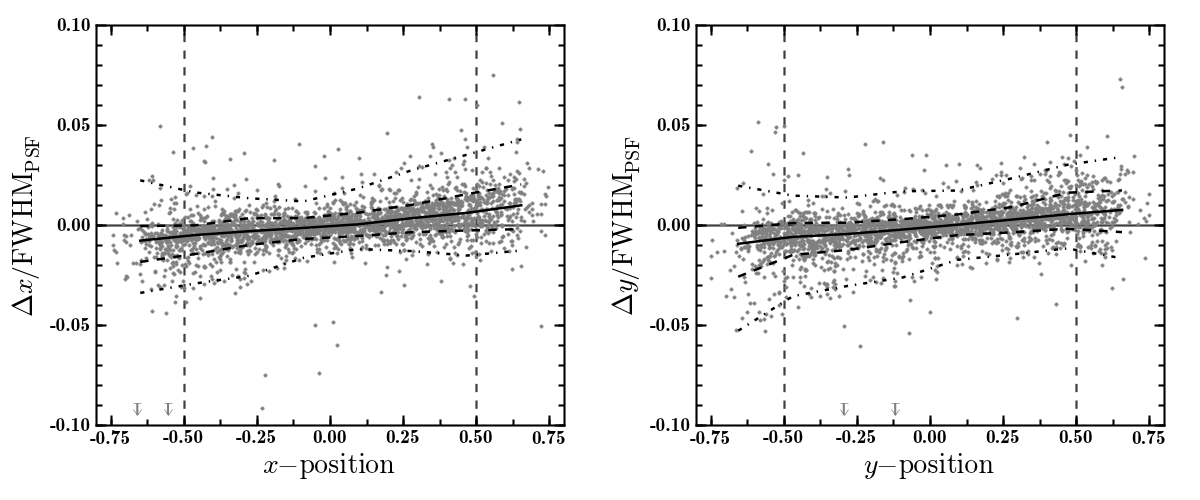}
 \caption{Comparison between the recovered and true source coordinates for the sources in 100 simulated crowded field datacubes, expressed as the mean deviation in $x$- and $y$-direction relative to the FWHM of the PSF as a function of the position of each source, in units of the FoV size.
Dashed vertical lines indicate the FoV edges. The median deviation along $x$ and $y$ is marked by a solid black line, the 68.3\% and 95.4\% percentiles of both distributions are indicated by the dotted and dashed black lines, respectively. A few outliers are out of the plotted range and indicated by upper limits.
}
 \label{fig:pos_recovery_xy}
\end{figure*}

We assume that a source catalogue based on high-resolution imaging already exists. The task is thus to find a global transformation from a reference system (i.e. the source coordinates in the catalogue) to the coordinate system of the datacube. At least four parameters are required to describe this transformation: a rotation angle $\alpha$, a pixel scaling factor $\xi$ and a shift along both spatial axes, $C$ and $D$. If we denote the coordinates of an object in the reference system by ($u^n$, $v^n$), we can write down the coordinate transformation for a single layer as
\begin{align} 
 x_k^n & = \xi_k \left( \cos\alpha_k\,u^n + \sin\alpha_k\,v^n \right) + C_k\, , 
 \label{eq:coordtransX_rot} \\
 y_k^n & = \xi_k \left( -\sin\alpha_k\,u^n + \cos\alpha_k\,v^n \right) + D_k\, .
 \label{eq:coordtransY_rot}
\end{align}
Substituting $A_k=\xi_k\cos\alpha_k$ and $B_k=\xi_k\sin\alpha_k$, Eqs.~\ref{eq:coordtransX_rot} and \ref{eq:coordtransY_rot} can be rewritten as
\begin{align} 
 x_k^n = A_k\,u^n + B_k\,v^n + C_k \,  ,  
 \label{eq:coordtransX} \\
 y_k^n = A_k\,v^n - B_k\,u^n + D_k \, . 
 \label{eq:coordtransY}
\end{align}
Eqs.~\ref{eq:coordtransX} and \ref{eq:coordtransY} define a system of $2N$ linear equations, where $N$ is the number of sources taken into account. While the transformation will be wavelength dependent, this dependency can be further constrained taking into account that:

(i) for instruments such as PMAS or ARGUS, where each spaxel is coupled to an optical fibre guiding the light into the spectrograph, neither $\alpha$ nor $\xi$ should depend on wavelength and

(ii) the variation of both $C$ and $D$ with wavelength due to atmospheric dispersion can be predicted using a model for the wavelength dependent atmospheric refractive index during the observation. Such predictions as a function of airmass and parallactic angle have been derived, e.g., by \citet{1982PASP...94..715F}; also see \citet{2012SPIESandinC} for an overview of different approaches. For the range of airmasses ($\lesssim2$) of our PMAS data, an offset up to $0\farcs2$ (equivalent to $0.4$ spaxels) is predicted across the covered wavelength range, quite similar to the values we actually measure.

One might use this information to eliminate the wavelength dependency in Eqs.~\ref{eq:coordtransX} and \ref{eq:coordtransY} already in the data reduction and then assume a single transformation for all layers of the cube. However, such a correction would involve resampling the data which we believe is to be avoided as much as possible. Instead, we start by finding a best-fit solution for every layer separately and then use the information to constrain the transformation a posteriori: After all layers have been analysed, we determine the IFU coordinates $x^n$ and $y^n$ of every source as a smooth function of wavelength. To fix the rotation angle and the pixel scaling to a common value for all layers of the datacube, the polynomial fits can be coupled in a way that the change in $x^n$ and $y^n$ with wavelength is the same for all sources $n$. Furthermore, that slope can be fixed to the one predicted by atmospheric dispersion.
This approach is very flexible with respect to how much a priori information should be used in determining the coordinates (e.g., the determination of the refractive index depends on an accurate knowledge of atmospheric properties such as humidity or air pressure that is not always available) and it uses the information from all layers together to obtain a final result.

Note that instead of fitting the parameters of Eqs.~\ref{eq:coordtransX} and \ref{eq:coordtransY}, we directly fit a low order polynomial to each $x^n$ and $y^n$. This is done because the variation of $A$ to $D$ with wavelength is not completely independent from one another. For example, to a certain extent a change in the scale factor $\xi$ (and thus $A$ and $B$) might be compensated by changes in the shifts $C$ and $D$, especially when the S/N in a layer is low. Such correlated variations of the parameters on small wavelength scales would be washed out by a polynomial fit, causing larger errors in the recovered IFU coordinates.

In this simple linear model, Eqs.~\ref{eq:coordtransX} and \ref{eq:coordtransY} do not account for distortions of the FoV. Given the small number of spatial pixels provided by most IFUs, this approximation should always be valid. If that should be not the case, a solution would be to expand the coordinate transformation to also include higher order terms of $u^n$ and $v^n$.
This will likely become important for more complex instruments such as MUSE (cf. Sect~\ref{sec:muse}), where the FoV is split into 24 different parts that follow independent optical paths through the instrument before finally getting dispersed by different spectrographs.

One important feature in our procedure is that the global transformation model can include sources with centroids actually outside of the FoV. Of course the accuracy of the positions will decrease with increasing distance since the transformation itself can only be constrained inside the FoV, but we are mainly interested in sources close to the observed field which still can have an influence on the light distribution inside the FoV, especially in the case of bright stars.

Again, we used our 100 simulated crowded field datacubes to assess the accuracy of the recovered positions. In Fig.~\ref{fig:pos_recovery_xy} we plot the offsets between the input and the recovered source positions, in units of the PSF FWHM in each cube, so that the results provide a generic measure of the achieved accuracy relative to the achieved spatial resolution.

We find that the standard deviations of the recovered coordinates from the true ones are of the order of $1$--$2\%$ of the ${\rm FWHM}$ of the PSF. The highest accuracy is achieved in the centre of the FoV where the coordinate transformation is best constrained. The scatter increases only slightly towards the edges of the FoV, but there is also a small systematic offset in the sense that at small $x$- or $y$-coordinates, the recovered values are on average smaller than the true ones, while they are larger on average for large coordinates. With regard to the coordinate transformation, this implies that the scale factor of the transformation $\xi_k$ is recovered too large.
This behaviour can be attributed to the small FoV that we have to deal with. It leads to a significant fraction of sources that contribute to the observed flux distribution having their centres outside the FoV. In case of a small mismatch between the recovered and the true position, the strongest residuals will emerge in the region around those two positions. The more this region is pushed away from the observed field, the smaller its effect on $\chi^2$ will be. Thus $\chi^2(\xi)$ is asymmetric towards higher values of $\xi$, causing the observed trend.
One can obviously avoid this behaviour by making assumptions about the coordinate transformation and fixing the value of $\xi_i$. However, in this case the pixel scale of the integral field spectrograph must be known very precisely, to better than $0.01\arcsec$ to achieve a comparable accuracy. We emphasize that this trend is extremely small and that in many cases the uncertainties of the individual high resolution coordinates will already be higher.

\subsection{Extraction of spectra}
\label{sec:fluxes}

Once the PSF model and all source positions are known with sufficient accuracy, the least-square solution for the source spectra becomes a linear equation. We have to minimise
\begin{equation} \label{eq:linear}
 \chi^2 = |\vec{\hat A} \vec{a} -\vec{b}|^2 \, .
\end{equation}
We denote $\vec{\hat A}$ the `PSF matrix' because it contains the PSF of every source in the fit. $\vec{a}$ contains the object fluxes we aim at and $\vec{b}$ is the data. Note that, because of the 3-dimensional data structure, $\vec{a}$ and $\vec{b}$ are also 3-dimensional and $\vec{\hat A}$ is even 4-dimensional. We can then express the individual components of Eq.~\ref{eq:linear} as
\begin{align}  
  \label{eq:linear:A}
 A_{i,j,k}^n & = \frac{\psf_{i,j,k}^n}{\sigma_{i,j,k}} \, , \\
  \label{eq:linear:a}
  a^n & = f^n \, , \\
  \label{eq:linear:b}
 b_{i,j,k} & = \frac{\mathfrak{d}_{i,j,k}}{\sigma_{i,j,k}} \, .
\end{align}
If we want to solve Eq.~\ref{eq:linear} in a least-squares sense we have to properly take into account the uncertainty $\sigma_{i,j,k}$ of each data value. To achieve this, both the data and the PSF coefficients have to be divided by the uncertainties \citep{1992nrfa.book.....P}. It can easily be verified that substitution of Eqs.~\ref{eq:linear:A} to \ref{eq:linear:b} into Eq.~\ref{eq:linear} yields Eq.~\ref{eq:chi2}.

The solution to Eq.~\ref{eq:linear} is obtained separately in each layer of the datacube without any coupling of the fluxes of an individual star in adjacent layers. In this sense, it is similar to performing photometry on each star in each layer (with known PSF and source coordinates) and obtaining the spectrum of each star as the combination of its individual monochromatic fluxes. However, in contrast to common crowded field photometry, the monochromatic fluxes in one layer are obtained simultaneously for all stars instead of measuring star after star.

For each spectrum that is obtained by solving Eq.~\ref{eq:linear} we can estimate the S/N per layer via
\begin{equation}
 S/N_k = f_k\left(\sum_{i,j} \frac{\psf_{i,j,k}^2}{\sigma_{i,j,k}^2}\right)^{1/2} \, .
 \label{eq:signal_to_noise}
\end{equation}
Note that Eq.~\ref{eq:signal_to_noise} only takes into account the noise expected due to the pixel uncertainties $\sigma_{i,j,k}$ and therefore gives only an upper limit which does not include any noise contributions introduced by connecting the pixels in the analysis. As we will show below, the true S/N in a deblended spectrum can be significantly lower.

It is worth spending a few more words on the handling of uncertainties in the analysis. Throughout this paper we assume that the uncertainties are known and correct, i.e that the true variance in each pixel is $\sigma_{i,j,k}^2$. The reduction of integral field data usually requires at least one step of resampling the data onto a regular grid. Resampling always introduces covariances between neighbouring pixels, although they are usually neglected in the further analysis. Consequently, the provided variance tends to underestimate the true uncertainty of a pixel value. However, this has no effect on the optimization which is based on $\chi^2$ minimization. Only the interpretation of absolute $\chi^2$ values with respect to consistency between a model and the data may become doubtful. 

Since Eq.~\ref{eq:linear} describes a linear least squares problem, its solution does not require any starting values and can be directly obtained by matrix inversion. For a realistic number of a sources this is not a computationally intensive process, even if applied to all layers in a cube.
Furthermore, the computing time required for the solution of Eq.~\ref{eq:linear} can be very significantly decreased further by taking into account the fact that the contribution of any star is limited to the pixels in its vicinity. Each element $n$ of the PSF matrix $\vec{\hat A}$ therefore contains a large number of pixels with essentially zero values. For this reason, $\vec{\hat A}$ can be written as a sparse matrix, for which there are dedicated efficient algorithms available \citep[e.g.,][]{1982ACMTM...8...43P}. With these provisions it becomes possible to fit the fluxes of even several thousand sources in a MUSE datacube simultaneously (see Sect.~\ref{sec:muse} below).
The computation time required to analyse a datacube is almost exclusively determined by the time required to obtain the PSF and the coordinate transformation. The computation time scales linearly with the number of sources. When fitting $\sim20$ sources for an instrument comparable to PMAS, one iteration on a single layer takes around 10~s on a single CPU. For the typical number of iterations and layers in a cube, the total time required for a whole datacube is around 10--20 hours on a single CPU. Some parts of the code have already been parallelized, others will be in the future, so that the actual time required for the analysis can be strongly reduced.

\subsection{Construction of the source list}
\label{sec:selection}

When the input catalogue of sources in the field is constructed from high-resolution imaging, it will typically reach magnitudes where the S/N level in the IFS data is too low to produce meaningful spectra. We thus need to construct a subsample of stars whose spectra can possibly be deblended in an available datacube. By design, this subsample will contain stars over a large range in magnitudes and expected S/N levels. Even if we are interested mainly in the stars bright enough to yield a spectrum with a sufficient S/N for some analysis, we also need to account for the effects of blending with the (more numerous) fainter objects. There is a limit, however: When true source confusion sets in, the deblending process and the flux assignment to individual sources becomes to some extent arbitrary. In Sect.~\ref{sec:performance} we explore quantitatively by means of simulations where this limit is reached.

The decision of whether or not to include a particular source will depend on several criteria:
(i) the brightness of the source,
(ii) the distance to other nearby sources and their relative brightnesses,
(iii) the position of the source, in particular if it is located close to the edge of the FoV (or even outside, see below). Effectively, the first two criteria can be combined into a single one based on S/N, taking the degradation of S/N due to crowding into account. 

The practical sequence of constructing a source list is as follows. We first estimate a limiting magnitude where confusion becomes dominant, based on a global characterisation of the exposure, given its depth, resolution etc., in comparison with the input catalogue. We then select a preliminary source list on the condition that the source magnitudes are brighter than the confusion limit. For those sources we predict the continuum S/N using simulations as explained below, accounting for the overall effects of crowding as well as for the influence of close-by bright stars. The final source list is then based only on the expected S/N of the spectra.

In this process there will be stars with magnitudes brighter than the confusion limit that do not pass the second selection stage, given their proximity to a brighter star. Yet the influence of those stars has to  be taken into account in the analysis. In such cases we generate a modified PSF for the close-by bright star that approximates the contribution of its fainter companion using the broad-band magnitudes of the two stars and their distance. The extracted spectrum will then be a combined one, and will be flagged accordingly in the resulting catalogue of spectra.

We note that the selection using S/N ratios will also take care of picking the sources close to the edges of the FoV that contribute significantly to the observed data.
Eq.~\ref{eq:signal_to_noise} is a sum over all spatial pixels, the flux that enters in the S/N calculation is the fraction of flux that is recorded by the detector.
For a given position, this fraction is determined by the PSF. The further a source is located outside the FoV, the smaller this fraction, and thus the S/N will be.

\subsection{Treatment of background}
\label{sec:background}

So far we concentrated our discussion on the resolved sources. But the data will always contain some flux contribution from unresolved background components. One such contribution is the night sky, the intensity of which we assume to be constant over the field of view. Just like any other flux component that is spatially flat, it can be easily accounted for by expanding the PSF matrix $\vec{\hat A}$ by a component whose values are inversely proportional to the uncertainties $\sigma_{i,j,k}$

Of greater interest is another background component that is produced by stars below the confusion limit, i.e. those that are not included in our source list discussed in Sect.~\ref{sec:selection}.
This background will in general not be spatially flat, but instead the distribution of the stars will produce a grainy structure, known from the surface brightness fluctuations observed in nearby galaxies. Our approach allows us to actually model these fluctuations. The sources above the confusion limit provide us with a model of the PSF in the observation. Additionally, from the high-resolution imaging in combination with our model for the coordinate transformation discussed in Sect.~\ref{sec:positions}, we obtain precise IFU coordinates even for the sources below the confusion limit. Furthermore, we can use the photometry provided by the high-resolution imaging to estimate the relative brightness of the sources. With PSF, positions and relative brightnesses in hand, we can predict the relative brightness of the grainy background in every spatial pixel of the datacube. This prediction can then be included (again after division by $\sigma_{i,j,k}$) in the PSF matrix.

Another physical background component could be in the form of gaseous emission, e.g., from filaments in an \ion{H}{II} region. Obviously, such a component is not to  be expected in globular clusters but we want to provide a tool that is useful in any kind of crowded stellar field. We also mention again that the scientific focus might actually not lie on the resolved stars but the nebular emission: One might be interested in the kinematics or emission line ratios and wish to remove the contamination of the bright stars. In their work on extragalactic planetary nebulae, \citet{2004ApJ...603..531R} showed that integral field spectroscopy is capable of separating individual point sources from diffuse emission of the ISM. The ISM component will be line emission and thus be restricted to a few layers in the datacube. So even if it strongly biases the PSF fit in those layers, a reliable PSF model can be obtained by interpolation of the results obtained bluewards and redwards of the emission line. One might then start from some initial guess for the spatial intensity distribution of the gaseous emission that is included in Eq.~\ref{eq:linear} and iteratively improve it based on the residuals observed in the layers under consideration.

\section{Performance of the deblending process}
\label{sec:performance}

The quality of the spectrum extracted from a datacube containing a single isolated star will depend almost exclusively on the brightness of that star. This is different for the extraction from crowded fields, where several effects may contribute to degrade the spectrum. We investigated the performance of our deblending and extraction algorithm on the basis of our simulated data\-cubes. We employed several criteria to quantitatively measure the quality of the extracted spectra:

The first two criteria are purely formal indicators: (i) We measured how the S/N in the continuum behaves under crowding conditions, relative to the value in isolated stars of the same magnitude. (ii) We also tested how robustly the continuum level is recovered, in terms of a potential systematic error in the broad band magnitude. 
 
In order to facilitate a discussion of the astrophysical possibilities and limitations of crowded field spectroscopy, we also considered the behaviour of two types of derived spectral parameters on crowding: (iii) radial velocities, and (iv) equivalent widths of strong absorption lines.

In the following we illustrate and discuss the performance of our code for each of these parameters. We first illustrate and quantify crowding effects for the restricted scenario of only two stars within a cube, with varying angular distance and brightness difference. We then explore the global performance in realistic situations using the `crowded field datacubes'  representing mock PMAS observations of a real globular cluster, constructed as described in Sect.~\ref{sec:sim}.

\subsection{``Crowding'' of two stars}
\label{sec:twostar_tests}

\subsubsection{Signal to noise ratio and continuum level}

\begin{figure}[tb]
 \resizebox{\hsize}{!}{\includegraphics{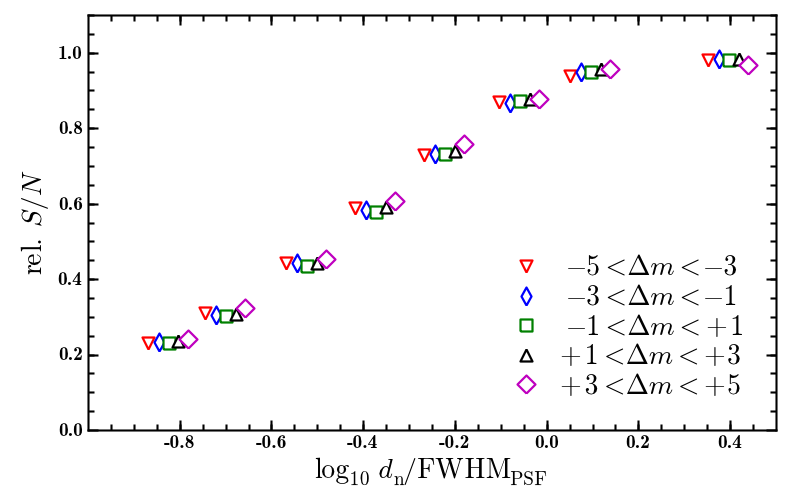}}
 \caption{
Degradation of the S/N ratio in the spectrum of a star with a nearby neighbour, as a function of the distance to the neighbouring source (normalized to the width of the PSF) and of the magnitude difference (star minus neighbour) between the two stars. We define the relative S/N as the ratio between the measured value and the value expected for an isolated star. Note that for clarity, small horizontal offsets have been applied to the plotted datapoints for the different magnitude differences.
}
 \label{fig:sim_catrip_twostar_snr}
\end{figure}

\begin{figure}[tb]
  \resizebox{\hsize}{!}{\includegraphics{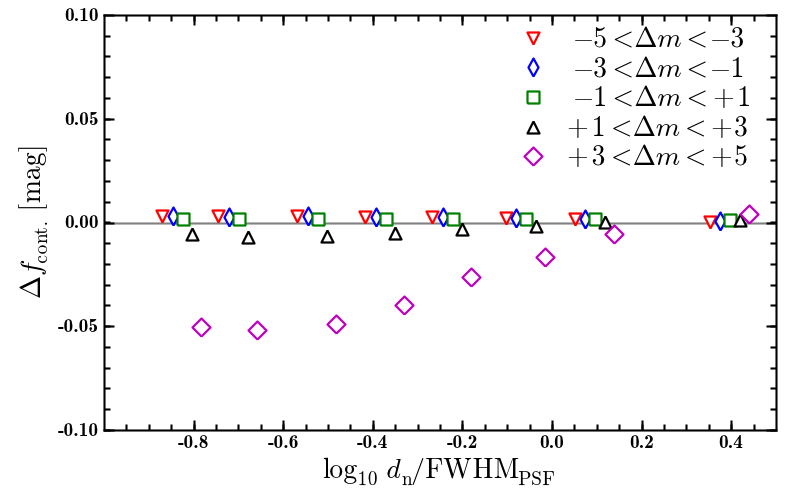}}
  \caption{
The average error in the continuum level of the deblended spectra (expressed as a magnitude difference) as a function of the distance of the source to a neighbour (again normalized to the width of the PSF) and its relative brightness. As in Fig.~\ref{fig:sim_catrip_twostar_snr}, datapoints for different flux ratios are plotted with small horizontal offsets for clarity.
}
  \label{fig:sim_catrip_twostar_cont}
\end{figure}

To assess how the S/N is affected by crowding, we compared the S/N expected using Eq.~\ref{eq:signal_to_noise} to the one actually measured in the deblended spectra.
To measure the S/N of a deblended spectrum we did the following: we subtracted the (noise free) input spectrum from the deblended one. To account for possible mismatches in the continuum level and slope of the extracted spectrum we fitted the residuals with a low order polynomial that was then also subtracted. The residuals should now scatter around zero with a standard deviation equal to the noise level of the deblended spectrum. We determined the standard deviation in a window around the central wavelength and divided the mean flux of the deblended spectrum in that window by it to yield a S/N ratio.

For two stars of varying angular separation and magnitude difference, Fig.~\ref{fig:sim_catrip_twostar_snr} shows the dependence of the S/N degradation on source separation $d_\mathrm{n}$ and flux ratio. For separations greater than the FWHM of the PSF, no degradation occurs as expected. For smaller separations the S/N decreases steadily and is only half that of an isolated star for $d_{12} \simeq 0.3$\,FWHM. The reason for this degradation is an increasing degeneracy in Eq.~\ref{eq:linear:A}. As the locations of the two stars approach each other, their PSF images become nearly the same, and the solution of Eq.~\ref{eq:linear:A} is no longer unique; any linear combination of the two stars that maintains the total flux will yield an almost equally likely (in terms of $\chi^2$) result. Practically this means that in every layer a different amount of flux will be transferred from one star to the other.

An interesting aspect revealed by Fig.~\ref{fig:sim_catrip_twostar_snr} is that the decline in S/N seems to be almost independent of the flux ratio between the two stars. This is unexpected at first sight -- one might expect a bright companion to have a higher impact on the deblended spectrum of a faint star than vice versa. But in fact the observed behaviour follows directly from Eq.~\ref{eq:signal_to_noise}: For small separations, the expected noise (the inverse of the square root in Eq.~\ref{eq:signal_to_noise}) will be almost the same for the two sources. But the final noise will also be the same for both sources because it is determined by the amount of flux that is shuffled back and forth between the two. Of course, the \emph{absolute} S/N will still be higher for the brighter source because its signal is higher.

It is interesting to view the results depicted in Fig.~\ref{fig:sim_catrip_twostar_snr} in the light of the Rayleigh criterion, which states that two point sources are resolved if their separation is larger than the FWHM of the PSF. It is of course well known that this criterion is not a strict limit but only indicative for the spatial resolving power. Fig.~\ref{fig:sim_catrip_twostar_snr} shows that in the case where the relative positions of the two sources are known, it is possible to deblend stars with distances well below the FWHM of the PSF, although at the price of a reduced S/N.

We also investigated how well the continuum level of the stars could be recovered. To this aim we characterise the continuum by dividing each deblended spectrum by the corresponding input spectrum and converting the result into a magnitude. In Fig.~\ref{fig:sim_catrip_twostar_cont} the dependence of the average continuum error on source separation and flux ratio is shown. It is remarkable that the strong decrease in S/N visible in Fig.~\ref{fig:sim_catrip_twostar_snr} does not cause a similar degradation of the continuum level. This implies that the increase in noise is indeed purely random and does not introduce any systematics in the deblended spectra. The fact that the actual deblending of the spectra (after PSF and source positions have been determined) does not require initial guesses makes it very robust against systematics. Especially in such cases discussed above where the two stars are very close to one another and the $\chi^2$ value becomes insensitive to the flux ratio between the two stars, the outcome of a fit that requires an initial guess would strongly depend on the value of that guess.

We do observe a small overestimation of the continuum level in the case of a very bright neighbour. However, the systematic error stays below $0.1\text{mag}$ and one should keep in mind that in this cases we are trying to measure the flux of a star that has a companion well inside the extent of the PSF that is at least $15\times$ brighter.

\subsubsection{Radial velocities and equivalent widths}

\begin{figure}[tb]
 \resizebox{\hsize}{!}{\includegraphics{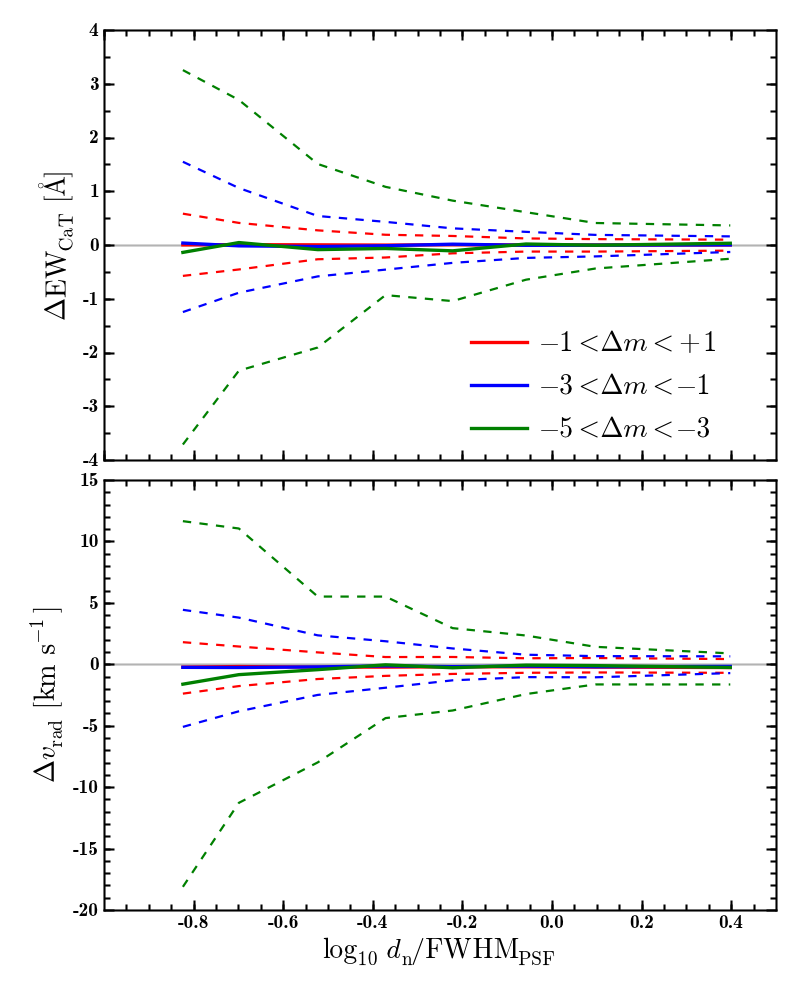}}
 \caption{
Accuracy in the recovered radial velocities (\textit{bottom}) and equivalent widths of strong absorption lines (\textit{top}) for a blended star, as a function of source separation and for different flux ratios. The lines depict the median velocity difference (measured $-$ true, thick solid) and the 75\% percentiles (thin dashed) for a distribution of 1000 stellar pairs per separation value.
}
 \label{fig:sim_catrip_twostar_phys}
\end{figure}

We now consider the recovery of astrophysical quantities. We determined radial velocities by cross-correlating the extracted spectra with the noise-free input spectra used to generate the datacube. The quantity of interest is then the 'measured minus true' velocity difference $\Delta v$. Recall that the spectral resolution adopted in the simulations was $\lambda/\Delta\lambda = 7000$, thus corresponding to a velocity resolution of 42.9 km/s (FWHM).   

The outcome of running our deblending code on a large number of simulated `two star datacubes' (1000 realisations per separation setup) is depicted in the lower panel of Fig.~\ref{fig:sim_catrip_twostar_phys}. The scatter of the velocity differences increases as the source separation decreases, which is of course expected as an immediate consequence of the declining S/N. However, the median values of $\Delta v$ stay comfortingly close to zero, although the two stars have randomly assigned velocity differences of the order of 10~km/s (see Sect.~\ref{sec:sim}). This is even so if the neighbouring star is much brighter than the analysed source.

But Fig.~\ref{fig:sim_catrip_twostar_phys} shows also clearly that the impact of partial blending in terms of the scatter on $\Delta v$ is strongest if the nearby star is significantly brighter. At first sight this behaviour might be seen as different from what we observe for the S/N (cf. Fig.~\ref{fig:sim_catrip_twostar_snr}), but it can be easily explained by the fact that the accuracy in the radial velocity measurement depends on the actual S/N, rather than on the S/N degradation.

To check whether the deblending process might lead to a systematic error in $\Delta v$ we divided our sample of stars into two subsamples, one with all stars that were assigned a positive `true' velocity relative with respect to their neighbours and one with all stars with negative `true' velocity. For both subsamples we found that the recovered median $\Delta v$ values are statistically indistinguishable from zero.

To investigate the behaviour of absorption line equivalent widths we focused on the Calcium triplet at $\lambda\lambda$8498, 8542, 8662~\AA. The equivalent width ($\ew$) of this feature is widely used to estimate the metallicity of stars \citep[e.g.,][]{2008MNRAS.383..183B}, and it is therefore important to look into the integrity of this quantity under crowding conditions.

The behaviour that we observe for the recovered values of $\ew$ is very similar to that just described for the radial velocities. The upper panel of Figure~\ref{fig:sim_catrip_twostar_phys} shows the median deviation and 75\% percentiles between the equivalent widths measured in the deblended spectra and those measured in the input spectra. On average there is again no bias, but the scatter increases with decreasing separation and with increasing brightness of the neighbouring star, as expected.

\subsection{Performance in realistic crowded fields}

In the last section, we showed how we can predict the expected S/N of a deblended spectrum under crowding (cf. Fig.~\ref{fig:sim_catrip_twostar_snr}), using the idealized case of only two stars. But when analysing a crowded stellar field, we have to take also another effect into account. Below a certain magnitude, the \textit{confusion limit}, the stellar density of similarly bright stars will be so high that they form a `pseudo background'. When this limit is reached, longer integration times will not lead to an increase of the number of resolved sources, though the average S/N will still increase.

In order to facilitate the following discussion, we first introduce the term \textit{resolution element} as the area covered by a circle whose diameter is equal to the FWHM of the PSF. When dealing with source densities, it is quite useful to specify them as numbers of sources per resolution element because this measure is independent of the specific instrument characteristics (number of spaxels, spaxel size) and observing conditions (seeing). Note that when stating source densities in the following, we refer to the density of stars brighter than a given limit.

Imaging studies are usually considered fairly complete down to source densities of 0.1 stars per resolution element. Of course, there is no sharp cut between detected and undetected sources at this limit as some brighter sources will already remain undetected while some fainter ones will still be found. Our analysis is based on an existing inventory of sources, so there is no need for a source detection. Instead, we define a \textit{resolvable} source as one that still improves the overall quality of the deblending process when it is included. Later, we will also discuss the subset of \textit{useful resolvable} sources, which are those for which physical parameters can be recovered to a given accuracy.

The aims we pursue in this section are twofold. First, we want to obtain a well-founded determination of the confusion limit, i.e., the transition from resolvable sources to unresolvable sources, and investigate the effects of selecting either too few or too many stars. This is an important aspect for the source selection process that was presented in Sect.~\ref{sec:selection}. Second, we want to verify whether the effects we discussed for the crowding of two stars (cf. Sect~\ref{sec:twostar_tests}) can also be identified in realistic crowded stellar fields.

We used the crowded field datacubes and tested how the number of stars included in the deblending process influences the results by including all stars in the deblending process brighter than a limiting magnitude $m_\mathrm{cut}$. $m_\mathrm{cut}$ was varied over a range that corresponded to average source densities between 5 and >100 stars per simulated datacube. In the following discussion, we will use two measures for the source density: besides the number of sources per resolution element we also give the number of stars per datacube, since absolute numbers are quite intuitive. In our application cases, one crowded field datacube contains 256 spatial pixels (reproducing the characteristics of the PMAS instrument) and because the FWHM of the PSF is 2 pixels, the number of resolution elements per cube is $\sim$80.

\subsubsection{Continuum biases}

\begin{figure}[tb]
 \centering
  \includegraphics[width=\hsize]{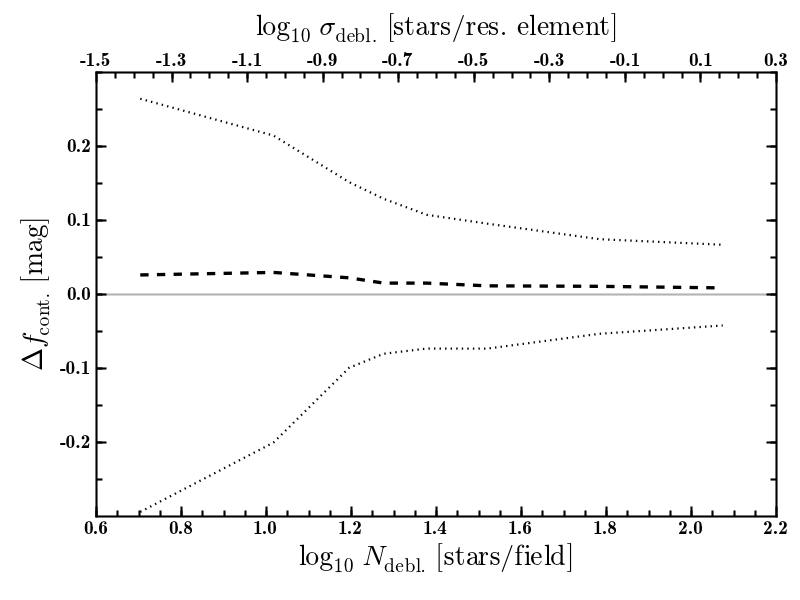}
  \caption[]{
The recovered continuum level of extracted spectra as a function of the number of stars in the source list, from the analysis of 100 simulated crowded field datacubes. The number of stars is given per datacube in the bottom and per resolution element in the top label. A thick dashed line is used to indicate the median continuum error of the brightest $10$ stars per cube, and dotted lines enclose the 75\% percentiles of the distribution. 

}
  \label{fig:sim_catrip_crowded_depth_cont}
\end{figure}

We first checked whether the continuum level of the stellar spectra becomes biased after extraction from a crowded stellar field. Again we converted the fraction between recovered and true spectrum into a magnitude. So in the case of systematic flux transfer to or from other stars due to source confusion, we expect a non-zero offset $\Delta f$. We performed the deblending experiment on 100 simulated datacubes and measured the distribution of $\Delta f$ for the set of the $10$ brightest stars in each datacube, as a function of the total number of stars in the source list. Recall that for each cube, the \emph{total} number of stars in the source list was varied from a few to $> 100$ per field. The results are based on the $10$ brightest stars per cube to allow for a fair comparison between the individual simulations. Using all deblended stars would penalize the simulations in which many stars were included because stars with fainter magnitudes will on average have deblended a spectrum with a lower S/N.

Figure~\ref{fig:sim_catrip_crowded_depth_cont} shows the median and 75\% percentiles of $\Delta f$ and its dependence on the number of stars in the source list. The median value of $\Delta f$ is nearly always very close to zero, implying that on average, the continuum level remains essentially unbiased under these crowding conditions. Only a very minor systematic offset to fainter magnitudes of $\sim0.01\text{mag}$ is observed. We expect that the missing flux is lost in the wings of the PSF and transferred into the background as it would agree with the fact that the PSF is recovered with slightly less pronounced wings (cf. Fig.~\ref{fig:psf_accuracy_statistics}).

It also becomes clear from Fig.~\ref{fig:sim_catrip_crowded_depth_cont} that individual stars can show significant deviations in their continuum levels. The accuracy in the recovered continuum is actually worst when only the few brightest stars ($<10$ stars per simulated crowded field datacube) are extracted because in that case stars falling just below the selection cut are still well resolved, yet their contribution is not accounted for during the deblending. $m_\text{cut}$ is significantly brighter than the confusion limit and thus the stars are extracted against a very inhomogeneous `pseudo'-background. This behaviour is quite similar to that reported by \citet{2006A&A...455..943M} when performing multi-object spectroscopy of horizontal branch stars in the globular cluster NGC6388: the contribution of close-by stars could not be accounted for and, as suggested by the authors, did likely influence the results. Our results show that such problems are essentially avoided when applying PSF-fitting techniques on IFS datacubes. The recovery of the continuum level is significantly increased upon including more sources in the deblending process.

\subsubsection{S/N degradation}

\begin{figure}[tb]
 \centering
  \includegraphics[width=\hsize]{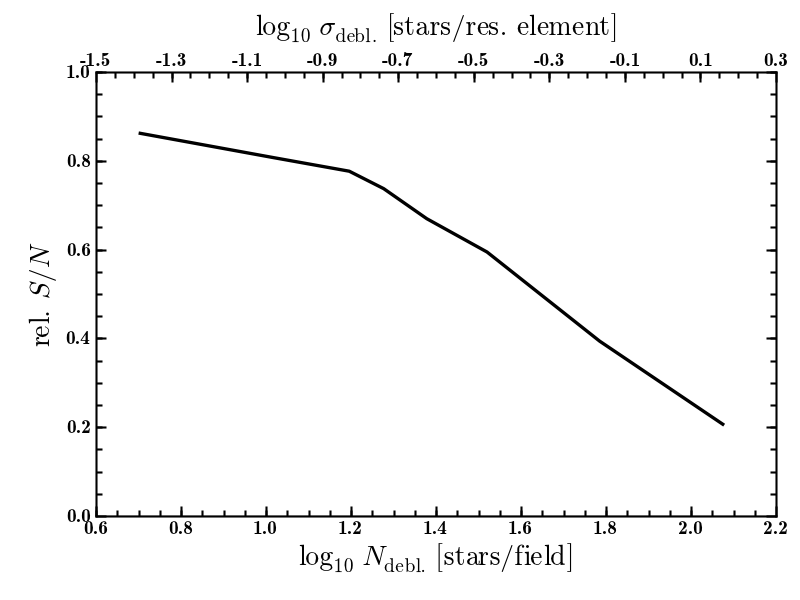}
  \caption{
Estimated S/N of extracted spectra as a function of the number of deblended stars per field or per resolution element, respectively, again using the $10$ brightest stars per simulated cube. The thick solid line indicates the median of the distribution. As in Fig.~\ref{fig:sim_catrip_twostar_snr}, we show the ratio of the S/N measured in a deblended spectrum and that expected based on Eq.~\ref{eq:signal_to_noise}.
}
  \label{fig:sim_catrip_crowded_depth_snr}
\end{figure}

The median relative S/N of the 10 brightest sources in each cube is shown in Fig.~\ref{fig:sim_catrip_crowded_depth_snr}. Apparently, the S/N in a given stellar spectrum decreases when the total number of sources taken into account is increased. This decrease can only be due to source confusion; in Sect.~\ref{sec:twostar_tests} we discussed the consequences of deblending sources with a small mutual distance.  Now, if we increase the number of stars in the process without taking into account the environment of a selected source, it becomes more likely that stars are included whose mutual distance is close to or even smaller than the minimum distance at which the spatial resolution of the data allows for a clean separation of the two. If one tries to deblend those sources nevertheless, this will lead to an increase in the noise of the extracted spectra. Yet the comparison with Fig.~\ref{fig:sim_catrip_crowded_depth_cont} also shows that this flux reshuffling is essentially random, individual stars on average do not receive flux or loose flux. Only their spectra get somewhat noisier. 

In the source selection scheme that we have adopted, such a behaviour is avoided by including the expected S/N in the selection process. For each source, the S/N is estimated using Eq.~\ref{eq:signal_to_noise} and applying the correction found in Fig.~\ref{fig:sim_catrip_twostar_snr}. Then only sources above a threshold in S/N are considered resolvable whilst others are added to a brighter neighbour.

\subsubsection{Recovery of spectral parameters}

\begin{figure}[tb]
 \centering
 \includegraphics[width=\hsize]{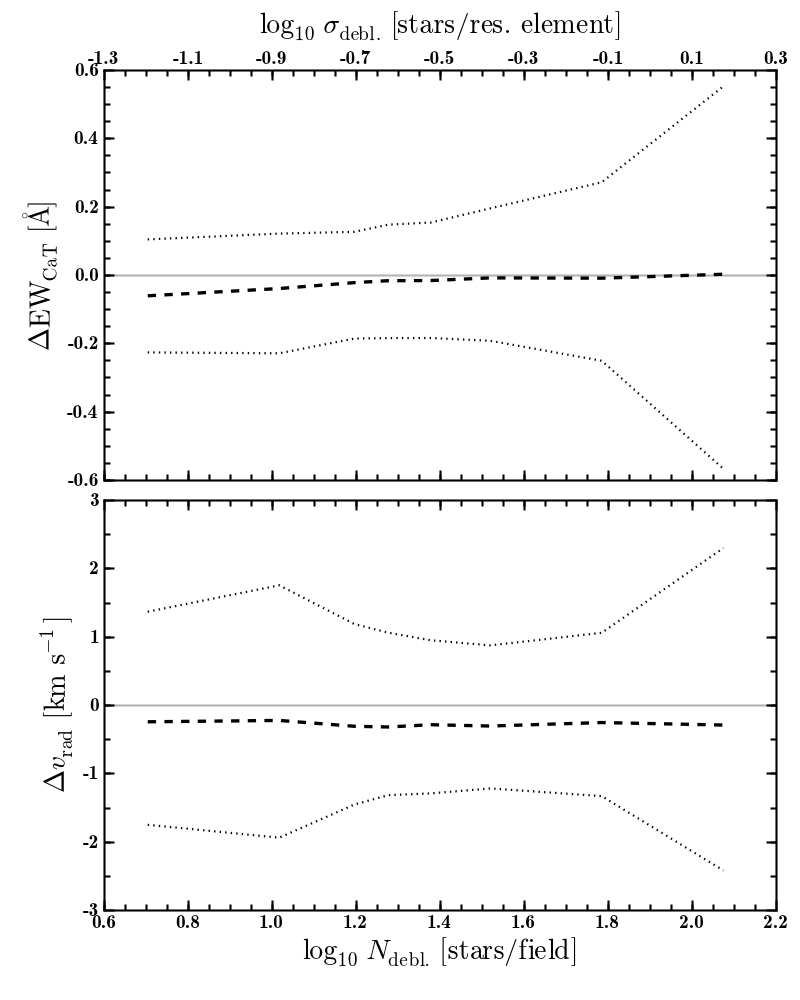}
  \caption{
Accuracy of the equivalent widths (\textit{top}) and radial velocities (\textit{bottom}) determined from the recovered spectra of the brightest $10$ stars per cube as a function of the number of sources included.
Line types are as in Fig.~\ref{fig:sim_catrip_crowded_depth_cont}.}
  \label{fig:sim_catrip_crowded_depth_line}
\end{figure}

To quantify the influence of the number of deblended sources on our ability to recover physical parameters from the spectra, Fig.~\ref{fig:sim_catrip_crowded_depth_line} shows the accuracy of measured radial velocities and equivalent widths as a function of the number of deblended sources.

When only the few brightest stars in each cube are deblended, the true $\ew$ are underestimated. This can be easily explained as a consequence of flux transfer from unaccounted fainter to brighter stars, because the $\ew$ increases as one moves up the red giant branch. Once the source list accounts for those fainter stars, this bias disappears entirely.

More interesting is the behaviour of the radial velocity accuracy $\Delta v$, which has a broad but clearly defined minimum. This suggests that there is actually an optimal number of stars to be used in the deblending process. The reason for the occurrence of such a minimum is that two counteracting effects influence the results, as we have identified previously. Inclusion of too few stars leads to systematic errors because the fluxes of fainter (yet still resolved) stars are not accounted for. Inclusion of too many stars on the other hand causes a drop in the S/N of the deblended spectra that renders the determination of stellar parameters less accurate. 

\begin{figure}[tb]
 \resizebox{\hsize}{!}{\includegraphics{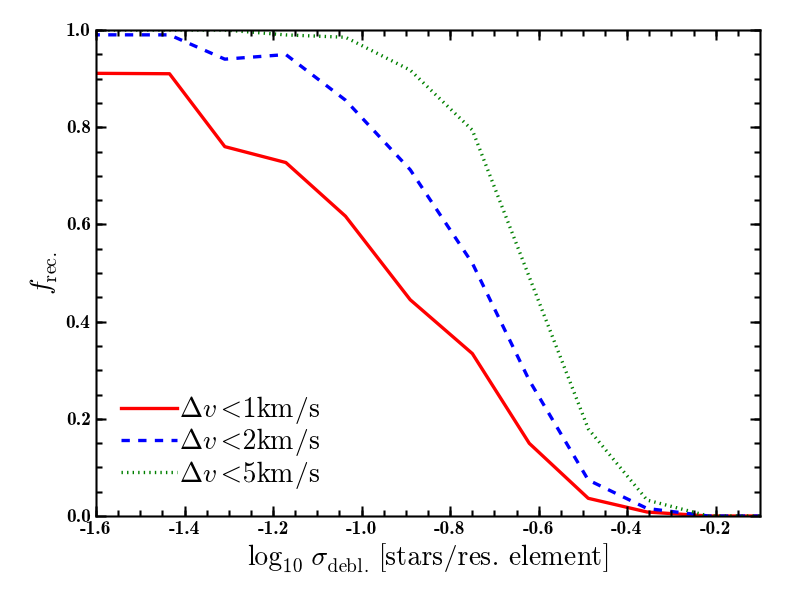}}
 \caption{The fraction of recovered sources is plotted as a function of the source density, using three different values for the required accuracy in the recovered radial velocity. The curves have been obtained using the same data that yielded the most accurate results in Fig.~\ref{fig:sim_catrip_crowded_depth_line}.}
 \label{fig:sim_catrip_binned_recovery}
\end{figure}

The source density at which those two effects are best balanced is $\sim$30 stars per simulated cube, corresponding to $0.4$ stars per resolution element. If we adopt this number as the confusion limit and compare to the value of $0.1$ stars per resolution element in crowded field photometry, we can quantify the improvement we get from using a pre-defined source catalogue instead of having to detect the sources in the datacube. Furthermore, we can use this value to give requirements on the quality of the input source catalogue in order to avoid being limited by the number of sources it contains: the stellar density of detected sources should be at least $4\times$ higher than the spatial resolution of the integral field data would allow for. Thus, the spatial resolution of the observation used to create the input catalogue must be at least twice the spatial resolution of the datacube.

So far, we have counted all stars as resolvable that yielded on average better results when included in the deblending process. Clearly, not every single star of those will yield a useful spectrum. The subsample of \textit{useful resolvable} stars will strongly depend on the science goals. To demonstrate this, we used the simulations that yielded the most accurate results, i.e., where $0.4$ stars per resolution element were deblended and determined the recovery fraction as a function of source density when requesting different levels of accuracy in the recovered radial velocities. The recovery fractions we obtained are shown in Fig.~\ref{fig:sim_catrip_binned_recovery}. For the very stringent condition that uncertainties are $\lesssim$1~km/s, the completeness drops to 50\% already at 0.1~sources per resolution limit. On the other hand, under more relaxed conditions, the completeness drops to below 50\% only at sources densities >0.2~sources per resolution limit.

Finally, we note that the recovered velocities have a very small systematic offset of $-0.2$~km/s, independently of the number of stars; for the present study this is however of no concern.

\subsection{Influence of crowding}
\label{sec:crowding}

\begin{figure*}[tb]
 \centering
 \includegraphics[width=17cm]{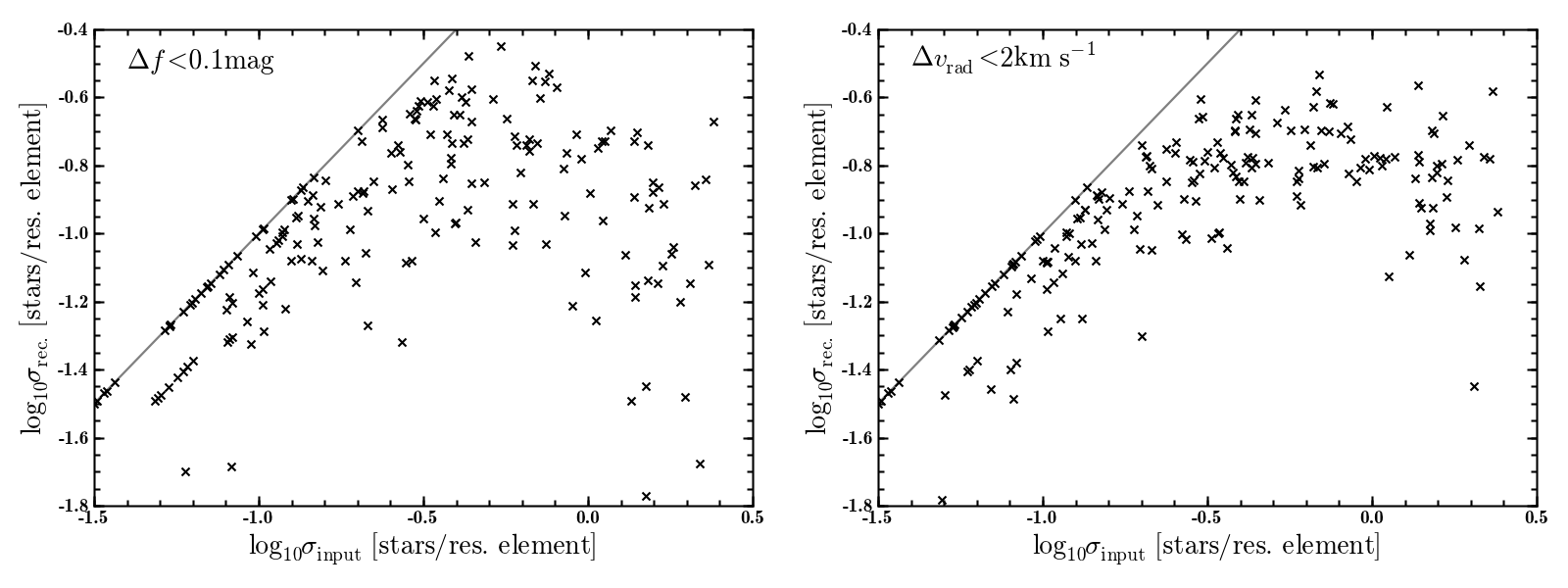}
 \caption{The impact of crowding on the efficiency of the deblending algorithm. We show the number of useful deblended spectra as a function of the number of bright sources in the simulation. To obtain a measure of the crowding that does not depend on the specific simulation setup (seeing, size of the field of view) we have normalized the number of sources by the number of resolution elements. Two criteria were used to identify a usefully deblended spectrum: a magnitude offset $<0.1\text{mag}$ (\textit{left}) and an offset in radial velocity $<2\text{km}\,\text{s}^{-1}$ (\textit{right}) between the input spectrum and the recovered one. Each plotted datapoint corresponds to the analysis of one mock datacube. In both panels, a grey solid line indicates the optimal case where a useful spectrum could be deblended for every bright star in the field of view.}
 \label{fig:sim_catrip_performance}
\end{figure*}

Our crowded field datacubes represent realistic integral field observations of a Globular Cluster, with stellar densities typical for Globular Clusters. We now want to investigate the performance of our method in different regimes of crowding. Over a certain range in stellar density we expect a trade-off between the crowding and the achievable depth: The more crowded the stellar field, the more our analysis will have to be restricted to the brightest stars. Yet with increasing stellar density, it will also get more challenging to deblend clean single object spectra at all because the contrast between the individual sources decreases. In the limiting case the stellar density will be so high already for the brightest sources that the stellar field is entirely unresolved.

In this section we aim to quantify up to what amount of crowding our approach yields useful results before it breaks down. Quantifying this limit is important when making predictions about whether a stellar field can be accessed by means of crowded field 3D spectroscopy. This will not only be applicable to Globular Clusters but to any type of crowded stellar field. In nearby galaxies, for instance, projected stellar densities can significantly exceed those of a typical Globular Cluster. Thus it would be very helpful to know up to what source density we can still obtain good results.

To test the influence of the crowding, we modified our simulations of crowded field datacubes in the following way: we identified as a bright star every source in the catalogue with a visible magnitude brighter than the Horizontal Branch ($\text{F606W}<13.5$ for 47Tuc), i.e we concentrated on the brightest giants. For each simulated datacube, we randomly picked stars from the catalogue and placed them in the datacube until a certain number of bright stars was reached. The number of bright stars picked varied between $4$ and $400$. The further processing of the cubes was then similar to the simulations described in Sect.~\ref{sec:sim}.

In total, $200$ datacubes were prepared that were all anal\-ysed using our algorithm. To quantify its performance for a single cube we counted the number of bright stars whose deblended spectra fulfilled an accuracy criterion. Two different accuracy criteria were used: an error in the recovered continuum of $<\!\!0.1\text{mag}$ and an offset in the recovered radial velocity of $<\!\!2\text{km}\,\text{s}^{-1}$. In Fig.~\ref{fig:sim_catrip_performance} we show the number of recovered sources as a function of the crowding. "Crowding" here is defined as the number of bright sources in a datacube. Furthermore, we again normalized the star counts by the number of resolution elements, for the reason mentioned above.

As Fig.~\ref{fig:sim_catrip_performance} shows, both accuracy criteria yield comparable results: We observe that up to a crowding of $0.2$ sources per resolution element, the number of accurately deblended sources increases approximately linear with the number of existing sources. For a higher crowding of $0.2$--$1.0$ sources per resolution element, we observe a plateau with an average of $\sim\!0.2$ accurately deblended sources per resolution element. If the crowding increases beyond $1$ source per resolution element, the number of deblended sources that fulfil our accuracy criteria starts to decrease again, indicating that we have reached an ``overcrowding regime'' where our approach breaks down.

We note that the distributions shown in Fig.~\ref{fig:sim_catrip_performance} can also be very helpful to judge whether single-star spectra can be deblended in a certain crowded stellar fields and to specify what spatial resolution would be required for its investigation.

\section{Potential sources for systematic errors}
\label{sec:systematics}

\subsection{Influence of the PSF}

\begin{figure}[tb]
 \resizebox{\hsize}{!}{\includegraphics{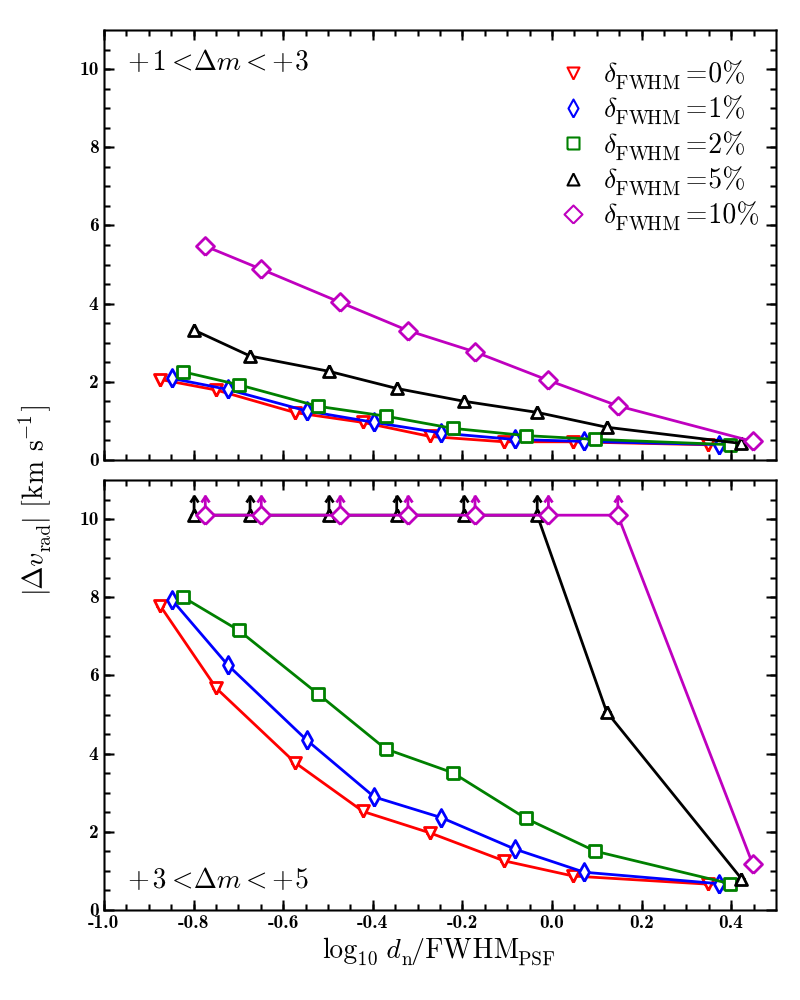}}
 \caption{
 Difference between recovered and true radial velocity of a simulated apparent binary star,
 as a function of source separation and of the degree of PSF mismatch, for two ranges of flux ratios between the two stars. The curves show the median of the \textit{absolute} velocity difference of each stellar pair. Also shown is the median offset for the case of a perfect PSF (red line) to show the behaviour expected in the absence of any systematic error. For clarity, deviations larger than the velocity dispersion have been set to a value of $10\text{km}\,\text{s}^{-1}$, the assumed velocity dispersion in the cluster.
}
 \label{fig:sim_catrip_twostar_psffwhm_vrad}
\end{figure}

In Sect.~\ref{sec:psf} we discussed to what accuracy the PSF can typically be determined in a crowded field datacube. We now investigate the effects of any possible mismatch between the true and the reconstructed PSF on the quality of the deblended spectra. To this aim, we modified the analysis of our two-star datacubes. The PSF assumed in the deblending process did now systematically deviate from the one that that was used to create the cubes. We describe the mismatch in terms of the FWHM of the PSF, as this is a quantity that is relatively easily accessible in observations. The analysis of Sect.~\ref{sec:psf} showed that we can recover the FWHM to an accuracy of usually $\sim\!1$--$2\%$, with some outliers at the $10\%$ level. Based on these results, we simulated PSF mismatches of $1$, $2$, $5$, and $10\%$ in FWHM.

Qualitatively, we expect the effect of an inaccurate PSF to be the following: The deblending process will leave residuals in the vicinity of every star. The amplitude of such residuals will scale with the brightness of the star. Close-by stars will then be deblended on top of the residuals and the extracted spectrum will be a combination of the true spectrum and the residuals. 

Quantitatively we wanted to know by how much these residuals bias a deblended spectrum and the derived astrophysical quantities. We measured again the radial velocities and Ca triplet equivalent widths in the deblended spectra, took the differences to the input values and checked the ensemble of results for systematic deviations. 

For obvious reasons the impact of any PSF mismatch will be strongest for relatively faint sources in the vicinity of significantly brighter ones. We therefore considered two cases: (i) A moderate brightness contrast between the source in question and its neighbour ($1 < \Delta m < 3$), and (ii) a strong brightness contrast of $3 < \Delta m < 5$. A contrast of 5~mag would be roughly the expected value for a star in the red clump of a Globular Cluster apparently close to a star at the tip of the red giant branch.

In Fig.~\ref{fig:sim_catrip_twostar_psffwhm_vrad}, we present the median absolute difference between the measured and true radial velocities as a function of the degree of PSF mismatch, in two panels corresponding to the different contrast classes. To discriminate between random and possible systematic errors, the median absolute difference is also shown for the case case of a perfect PSF (red line in Fig.~\ref{fig:sim_catrip_twostar_psffwhm_vrad}). In this case, the offset should be completely caused by random errors, i.e. the limited S/N of the deblended spectra. Any increase in the offset can then be attributed to the imperfect PSF. With increasing influence of the PSF residuals of the brighter neighbouring star, the median of the distribution will approach the assumed velocity dispersion. 

As expected, the influence of the PSF increases with the brightness contrast between source and neighbour. This can be verified by comparing with the two panels of Fig.~\ref{fig:sim_catrip_twostar_psffwhm_vrad}. For moderate brightness contrast, the introduced systematics are small if the FWHM of the PSF is determined to an accuracy of $<$5\%, whereas in the case of a strong brightness contrast, offsets of $2$\% already introduce measurable systematics. At this contrast level, the residuals caused by the PSF mismatch if the errors in the FWHM are $\geq$5\% are so strong that the signal of the fainter star basically disappears and no useful spectrum can be deblended any more. On the other hand, Fig.~\ref{fig:sim_catrip_twostar_psffwhm_vrad} also shows that PSF mismatch only becomes an issue for source separations comparable to or smaller than the FWHM. We find similar results regarding the accuracy of the recovered values of $\ew$.

Recall that for crowded fields in Galactic globular clusters we typically can recover the PSF width to an accuracy of $<$2\%. The simulations presented in this subsection demonstrate clearly that this will be sufficient to deblend an unbiased spectrum of a star in the close vicinity, to even a small fraction of the PSF width, of a neighbour that is $\sim$10$\times$ brighter. Only in the extreme case where the brightness contrast between the two stars is significantly larger than a factor of $10$, significant biases are to be expected. Yet such cases will be known from the input catalogue and can thus be easily flagged and excluded from the further analysis.

\subsection{Further sources for systematic errors}

An imperfect PSF is not the only potential source of systematic errors. Another such source are the positions of the individual stars in the field. For reasons like measurement errors or proper motions, the source coordinates that are given in the input catalogue can be offset from their true values. The strength of such effects is essentially a property of the specific input catalog that is used and cannot be predicted like the achieved accuracy of the PSF. For this reason, we did not try to quantify the influence of these effects on the deblended spectra.

A further error source related to the input catalogue is the issue of missing stars or spurious detections. Those effects will occur mainly in the regime of the fainter stars that are inaccessible to the integral-field observations (that have a lower spatial resolution). Yet under certain circumstances, they might also play a role amongst the brighter stars that are accessible. One such example are catalogues that are compiled from observations with the Advanced Camera for Surveys (ACS) onboard HST. In the case of globular clusters, those observations are usually targeted at the numerous faint main-sequence stars and the brightest giants appear heavily saturated in the exposures and cause strong bleeding features on the CCD that might cover relatively bright stars (see Fig~\ref{fig:muse_sim_fov} for an example). Again this effect will largely depend on the quality of the used input catalogue.

\section{MUSE}
\label{sec:muse}

\begin{figure}[tb]
 \resizebox{\hsize}{!}{\includegraphics{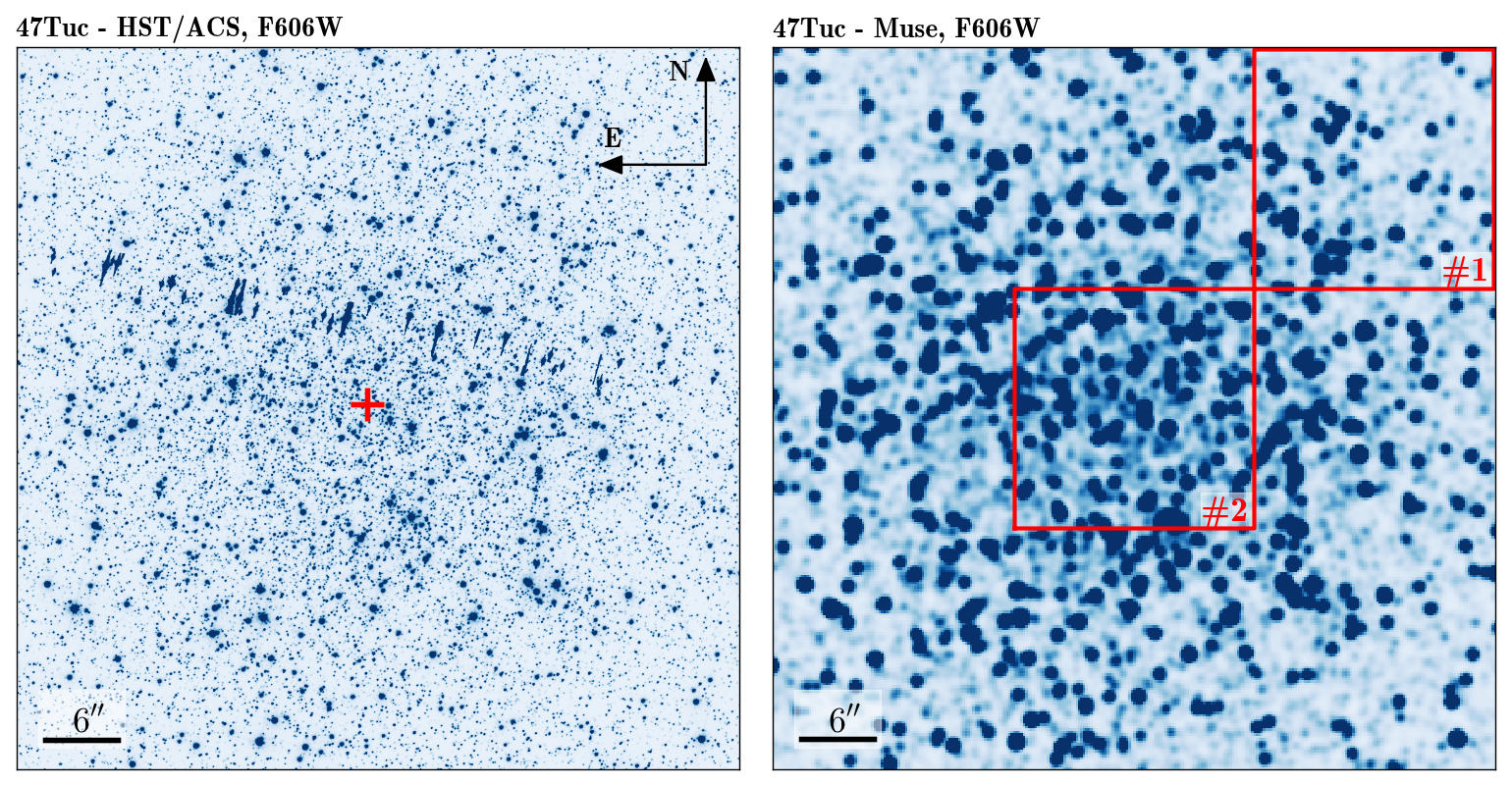}}
 \caption{
Simulated MUSE observation of the globular cluster 47Tuc.
\textit{Left:} Cut-out from an HST/ACS observation of the central arcmin of the cluster in the F606W-passband.
A red cross denotes the cluster centre.
\textit{Right:} Reconstructed broadband image from the mock MUSE data of the same region, obtained by integrating the datacube with the F606W filtercurve.
The seeing in the simulation was set to $0.8\text{arcsec}$.
Red squares indicate the two $20\times20\,\text{arcsec}$ fields that are discussed in the text.}
 \label{fig:muse_sim_fov}
\end{figure}

\begin{figure*}[tb]
 \centering\includegraphics[width=17cm,clip]{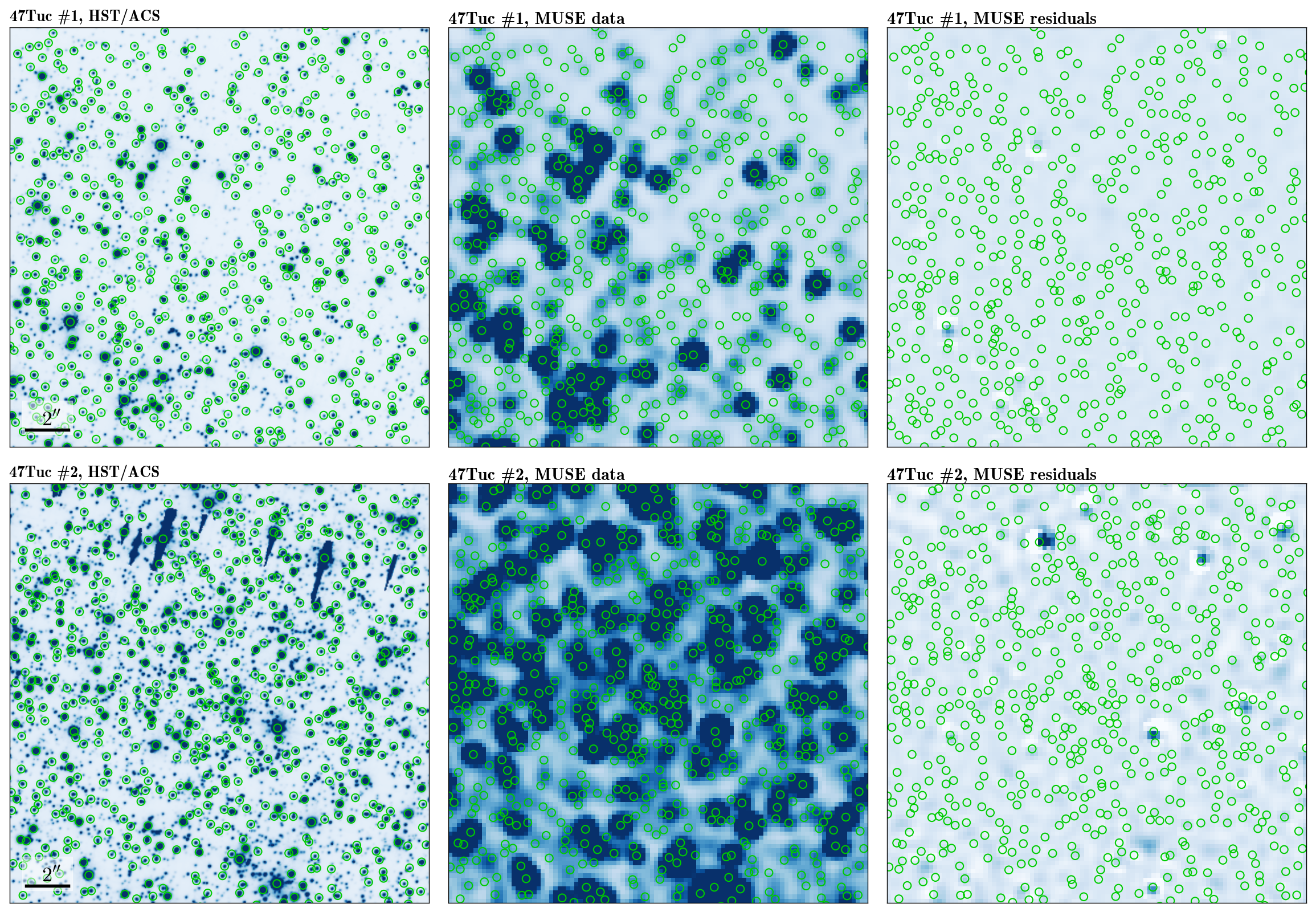}
 \caption{
Visualization of our source deblending algorithm applied to simulated MUSE data.
For each of the two fields highlighted in Fig.~\ref{fig:muse_sim_fov}, we show a cut-out from an HST/ACS image (\textit{left}), a white light image of the simulated data (\textit{centre}), and a residual image after fitting the sources (\textit{right}).
Each location where a useful spectrum was deblended has been marked by a blue circle.}
\label{fig:muse_sim_subfields}
\end{figure*}

MUSE \citep{2010SPIE.7735E...7B} is an integral field spectrograph currently being built  by a consortium of 6 European institutes and ESO. It is scheduled to see its first light at the Very Large Telescope (VLT) in 2013. The instrument provides a FoV of $1{\text{arcmin}}^2$ with spaxels of $0.2\times 0.2$~arcsec$^2$, and a wavelength coverage of $4650$--$9300{\text{\AA}}$. The combination of a large FoV with a spatial sampling sufficient to properly sample the PSF even under good seeing conditions makes MUSE a unique instrument for a variety of science applications. Although the main motivation for developing this new instrument is the observation of faint galaxies at medium to high redshift, some very promising applications exist for the investigation of crowded stellar fields. To demonstrate this, we outline in the following the analysis of a simulated MUSE datacube of the Globular Cluster 47Tuc.

The simulations are again based on the HST-photometry obtained in the HST/ACS Survey of Galactic Globular Clusters. Based on broadband colours and an isochrone fit to the colour magnitude diagram of 47Tuc, each star was assigned a spectrum based on a new library of model atmospheres and synthetic spectra calculated by Husser et al. (submitted) using the stellar atmosphere code PHOENIX \citep{1999JCoAM.109...41H}.

To simulate the effect of missing stars in the vicinity of brighter stars, we applied the following correction to the input catalogue: we counted the surface density of stars at a given magnitude in the vicinity of brighter stars and compared it to the overall density of those stars across the field covered by the catalogue. Stars were then randomly added in the vicinity of the brighter ones until the two densities matched.The catalogue that was later used in the analysis did not include those stars.

In the final step of the simulation, a datacube was created using dedicated software developed within the MUSE consortium (R. Bacon, private communication). It creates a datacube containing the provided sources and a sky spectrum. Each spectrum is convolved with the line spread function of MUSE. The seeing in the simulation was set to 0.8 arcsec. This value is internally translated to a wavelength dependent PSF. The final simulated datacube is a combination of three snapshot exposures, each with an exposure time of $30\text{s}$. Fig.~\ref{fig:muse_sim_fov} shows a whitelight image of this datacube together with an HST image.

We concentrate our discussion on two $20\times20\,\text{arcsec}$ subcubes, highlighted in Fig.~\ref{fig:muse_sim_fov} (\textit{right}) by red squares. In both regions, we deblended the stellar spectra using our crowded field spectroscopy code.

The algorithm itself was slightly modified in order to handle the significantly larger amount of data of a MUSE cube compared to PMAS or ARGUS. The size of the FoV of MUSE is sufficiently large so that some relatively isolated bright stars should exist within the FoV that can be used as PSF calibrators. We therefore used the existing photometry to select suitable PSF stars on the condition that within a given radius there are no neighbours brighter than a given flux ratio. The search radius is chosen such that the PSF contribution outside this radius is essentially zero. After each fit to the object fluxes during the iteration, we then subtract all stars except those previously identified and determine the coordinate transformation and the PSF using only those stars. This significantly speeds up the analysis and we achieve computational times similar to our PMAS or ARGUS datacubes. Nevertheless, the actual extraction of the spectra still is performed on \textit{all} stars simultaneously.

In Fig.~\ref{fig:muse_sim_subfields} we show a close-up of the two regions, again using an HST/ACS image and a whitelight image of the mock MUSE data. We also show a whitelight image for each region with the MUSE data where the deblended sources were already subtracted. Closer inspection of these residuals reveals that some stars have been missed by our source selection; these are the stars added to the incomplete source catalogues as discussed above. Such sources can be easily identified in the residuals and then added manually to the catalogue. 

To visualize the efficiency of our deblending approach we marked the position of every source for which a useful spectrum was deblended. Note that sources for which the extracted spectra have a S/N too low for a reliable radial velocity determination are not marked in Fig.~\ref{fig:muse_sim_subfields}. The total number of useful spectra that were deblended is $580$ in subfield \#1 and $610$ in subfield \#2. Interpolating these numbers to a full datacube, we estimate that from a single MUSE observation obtained under average seeing conditions we can obtain $\sim\!5000$ useful spectra; under very good seeing conditions this number may even be 3--4$\times$ higher.

\begin{figure}[tb]
 \resizebox{\hsize}{!}{\includegraphics{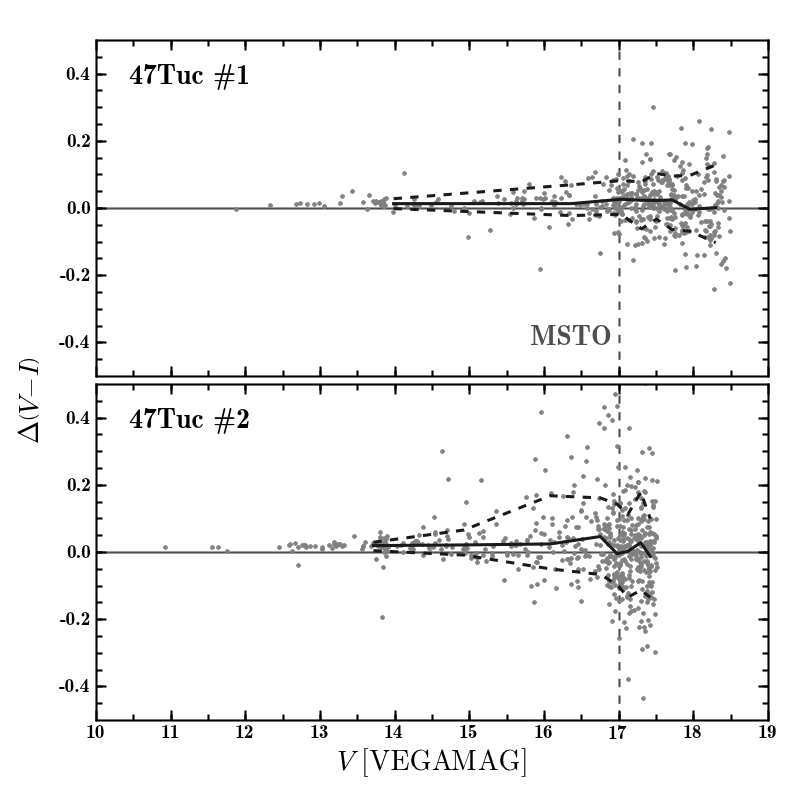}}
 \caption{
Deviation in $V$--$I$ colour recovered from the deblended spectra as a function of a stellar magnitudes for the two subfields highlighted in Fig.~\ref{fig:muse_sim_fov}. The median of the distribution is shown as a thick solid line, dashed lines give the $75\%$ percentiles of the distribution. A vertical dashed line is used to indicate the main-sequence turn-off in both panels.
}
 \label{fig:muse_sim_color}
\end{figure}

\begin{figure}[tb]
 \resizebox{\hsize}{!}{\includegraphics{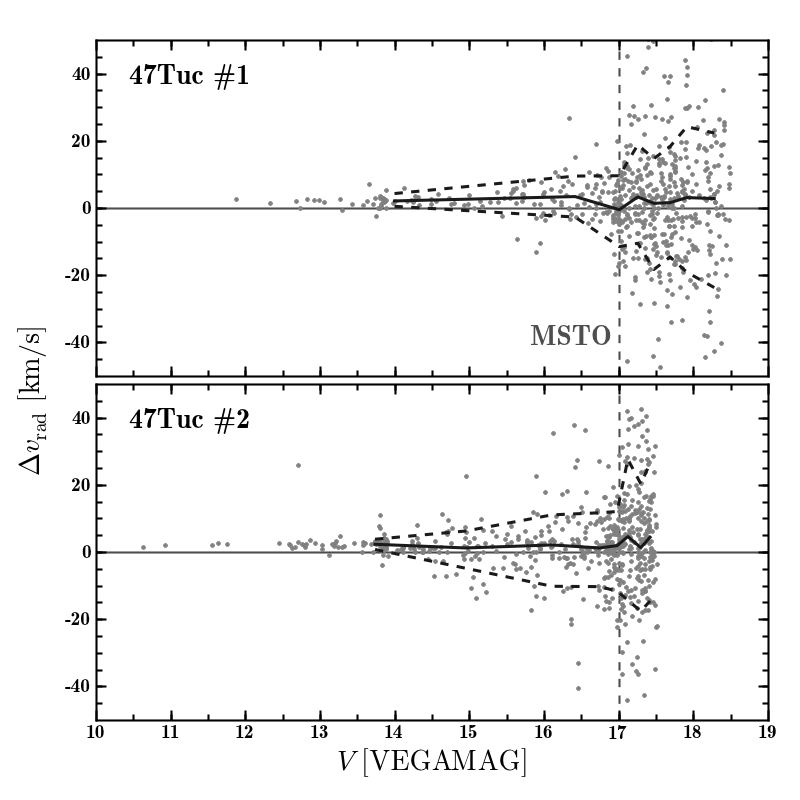}}
 \caption{
Deviation of the radial velocities determined from the deblended spectra, again plotted as a function of a stars brightness for the two fields highlighted in Fig.~\ref{fig:muse_sim_fov}. Radial velocities were determined by cross-correlating each deblended spectrum with its input spectrum. The line types are as in Fig.~\ref{fig:muse_sim_color}.}
 \label{fig:muse_sim_vrad}
\end{figure}

The extended wavelength range of MUSE allows us to directly determine broadband colours from the spectra by applying the filtercurves of the HST/ACS $F606W$ and $F814W$ filters (hereafter called $V$ and $I$, respectively) and compare them to the `true' (i.e., input) colours. In Fig.~\ref{fig:muse_sim_color}, the deviation in $V-I$ is plotted as a function of $V$ band magnitude. The comparison of the two fields gives a good impression of the effect of crowding: In the less crowded part of the simulated data (field \#1), we obtain useful spectra for fainter stars than in the direct vicinity of the cluster centre (field \#2). Yet in both fields we are able to probe below the main sequence turn-off. Taking into account that the simulation assumed only average seeing conditions and that MUSE is also designed to work with adaptive optics, this demonstrates the unique capabilities provided by the instrument.

Finally, we take a look at the accuracy achievable in the measured radial velocities. The spectral resolution provided by MUSE is smaller than what we used in our previous simulations and that of our existing PMAS and ARGUS data. However, the larger wavelength range at least partly compensates for this. As Fig.~\ref{fig:muse_sim_vrad} shows, radial velocities of bright giants can be determined with an accuracy of a few km/s. Around the main sequence turn-off of the cluster, the typical error of the obtained radial velocities is still comparable to the velocity dispersion.

\section{Conclusions}
\label{sec:conclusions}

The application of PSF fitting techniques to integral field spectroscopy is a powerful approach to observe crowded stellar fields. We developed an algorithm to deblend the spectra of many stars within a single datacube simultaneously, and validated the method by applying it to realistic simulated observations of the central regions of globular clusters. The combination of linear least-squares fitting with the usage of sparse matrices makes the code computationally efficient and affordable even for a modest workstation.

One central assumption for our algorithm is that an input catalogue of the sources in the field already exists. Typically this catalogue would be obtained by other means, such as by running a classical crowded field photometry code on high-resolution HST images. It is of course also possible to perform the source detection in the IFS datacubes themselves, e.g.\ from a collapsed white-light image. But our simulations show that by using prior knowledge of the locations of sources in the field, the number of correctly deblended sources is increased by up to a factor of $\sim$4 compared to the case where the source detection is performed on the IFU data. 

We have extensively tested the performance of our code as a function of the degree of crowding, expressed as the number of sources per field or per spatial resolution element. A conventional rule of thumb for crowded field \emph{photometry} states that deblending performs well up to a stellar density of $\sim$0.1 per resolution element. We have shown that the spectroscopic deblending works at even considerably higher source densities than that. This gain is partly due to the application of prior knowledge as discussed above, and partly due to the continuity enforcement over many simultaneously evaluated image layers in a datacube.

While unbiased spectra can be extracted even for heavily blended sources, such spectra will suffer from a significantly reduced S/N level, with the degradation being driven by the proximity to and the brightnesses of nearby stars. We have shown that this reduction of S/N due to blending can be accurately modelled and predicted for a given dataset from the input catalogue. Consequently, an optimal source list for the final extraction can be constructed according to the expected S/N of the final spectra. This is a very useful feature for statistical investigations in crowded stellar fields, as it allows one to maximise the number of `meaningful' spectra that can be obtained from a given dataset. 

Under conditions of strong crowding, the number of stars per field for which spectral parameters can be reliably determined is approximately independent of the actual source density and corresponds to roughly 0.2 stars per spatial resolution element. This `plateau' exists because of the mutually opposing effects of higher source densities on the one hand, and more severe S/N degradation due to crowding on the other hand. Only when the observed density of stars of comparable brightness passes a definite `overcrowding limit' of $\sim$1 star per resolution element, the extraction of useful individual spectra breaks down entirely.

The degree of crowding in a given field depends of course also on the depth and angular resolution of the data. We constructed our simulated datacubes in view of our own existing observations of Galactic globular clusters, using present-day IFUs under seeing-limited conditions.  Even in the very central regions of these clusters we found that the source density of stars bright enough to produce meaningful spectra was still well below the overcrowding limit. However, this could change rapidly if the data were going deeper down the colour-magnitude diagram, especially once the main sequence is being probed. We have demonstrated that with the upcoming MUSE instrument this domain will actually be reached. Improving the angular resolution, for example through ground-layer adaptive optics as envisaged for MUSE, will then become crucial. 

The spectroscopy of crowded stellar fields may be of interest for other classes of astronomical objects, such as compact open clusters, dwarf galaxies, or dense regions in the bulge of the Milky Way. The methodical work presented in this paper will enhance the capabilities of `crowded field 3D spectroscopy' beyond our own application topic of globular clusters. For the benefit of the community, we plan to make our code available to the public in the future. 

\begin{acknowledgements}
We thank the anonymous referee for a careful reading of the manuscript and many useful comments that helped to improve this paper. \newline{}

S.~K. acknowledges support from the ERASMUS-F project through funding from PT-DESY, grant no. 05A09BAA. \newline{}

The authors thank Eric Emsellem and Guiseppina Battaglia for providing a tool to facilitate the assignment of realistic spectra to the photometric data. We are grateful to Roland Bacon for sharing the QSim software to create realistic MUSE datacubes. Tim-Oliver Husser has been a great help in the preparation of the MUSE data.
\end{acknowledgements}

\bibliographystyle{aa}
\bibliography{kamann_cf3ds_arxiv_v1}

\end{document}